\documentclass[usenatbib]{mnras}
\usepackage[T1]{fontenc}
\usepackage{ae,aecompl}
\usepackage{amsmath}
\usepackage{amssymb}
\usepackage{mathtools}
\usepackage{bm}
\usepackage{cleveref}
\usepackage{graphicx}
\usepackage{caption}
\usepackage{subcaption}
\usepackage{float}
\usepackage{txfonts}
\usepackage{hyperref}
\usepackage{multirow}
\usepackage{hhline}
\title[Extra Galactic  GCs - Milky Way \& Andromeda]{MOCCA-Survey Database: Extra Galactic Globular Clusters. II.  Milky Way and Andromeda}

\author[A. Leveque et al.]{
A.~Leveque$^{1}$\thanks{E-mail:agostino@camk.edu.pl},
M.~Giersz$^{1}$,
M. Arca-Sedda$^{2}$ 
\& Abbas Askar$^{3}$
\\
$^{1}$ Nicolaus Copernicus Astronomical Center, Polish Academy of Sciences, ul. Bartycka 18, PL-00-716 Warsaw, Poland\\
$^{2}$Astronomisches Rechen-Institut, Zentrum f\"{u}r Astronomie der Universit \"{a}t Heidelberg, M\"{o}nchhofstr. 12-14, D-69120 Heidelberg, German\\
$^{3}$ Observatory, Department of Astronomy, and Theoretical Physics, Lund University, Box 43, SE-221 00 Lund, Sweden\\
}
\graphicspath{{Figures/}}

\begin{document} 

   \date{}
   \pagerange{\pageref{firstpage}--\pageref{lastpage}} \pubyear{2022}
   \maketitle

 
  \begin{abstract}
A comprehensive study of the co-evolution of globular cluster systems (GCS) in galaxies requires the ability to model both the large scale dynamics (0.01 - 10 kpc) regulating their orbital evolution, and the small scale dynamics (sub-pc - AU) regulating the internal dynamics of each globular cluster (GC). In this work we present a novel method that combine semi-analytic models of GCS with fully self-consistent Monte Carlo models to simultaneously evolve large GCSs. We use the population synthesis code MASinGa and the MOCCA-Survey Database I to create synthetic GC populations aimed at representing the observed features of GCs in the Milky Way (MW) and Andromeda (M31). Our procedure enables us to recover the spatial and mass distribution of GCs in such galaxies, and to constrain the amount of mass that GCs left either in the halo as dispersed debris, or in the galactic centre, where they can contribute to the formation of a nuclear star cluster (NSC) and can bring stellar and possibly intermediate mass black holes there. The final masses reported by our simulations are of a few order of magnitudes smaller than the observed values. These differences show that mass build-up of a NSC and central BHs in galaxies like MW and M31 cannot be solely explained by the infalling GC scenario. This build-up is likely to depend on the interplay between interactions and mergers of infalling GCs and gas. The latter can contribute to both \textit{in-situ} star formation in the NSC and growth of the central BH.
\end{abstract}

\begin{keywords}
globular clusters: general
\end{keywords}

%

\section{Introduction}
The advent of the HST and large ground based telescopes (for example E-ELT, SALT, LSST) increased significantly the level of detail of globular cluster (GCs) observations \citep[see \citealt{Brodie2006,Kruijssen2014,Renaud2020} and references therein]{Larsen2001,Cote2004,Peng2006,Peng2008}. One crucial discovery  has been the  bimodality of GCs in the color distribution, indicating two subpopulations of GCs around the host galaxy, with one peak shifted toward blue, indicating a metal-poor population, and the other red, indicating a metal-rich population. This feature seems to be common in all types of galaxies \citep{Zepf1993,Ostrov1993,Kundu2001,Larsen2001,Harris2006,Cantiello2007}. The two blue and red peak locations differ from galaxy to galaxy. However, the $V-I$ color distribution for bright early-type galaxies (like NGC 1023, NGC 3384, NGC 4472 for example) usually shows a blue peak at $V-I = 0.95 \pm 0.02$, corresponding to $[Fe/H] \sim -1.5$ and a red peak at $V-I = 1.18 \pm 0.04$, corresponding to $[Fe/H] \sim -0.5$ \citep{Larsen2001}. 
Despite different scenarios having been suggested to explain the observed color distribution (\citet{Ashman1992}; \citet{Forbes1997}; \citet{Cote1998}), no consensus has been reached. In the hierarchical scenario, among others, the metal-rich GCs would be created \textit{in-situ}, with the metal-poor GCs accreted from lower-mass galaxies to more massive galaxies \citep{Forbes1997,Harris1999,DAbrusco2016,Cantiello2018,Cantiello2020}. Instead, at least two star-formation events in the histories of such galaxies has to be invoked to generate such a bimodality, which can be triggered by major mergers \citep{Ashman1992} or occur in isolation \citep{Forbes1997}. 

As suggested by \cite{Harris2013}, the total number (or mass) in GCS in a galaxy seems to ubiquitously increase with the host galaxy mass, or luminosity. The authors found that the low and very high luminosity galaxies show a larger number of surrounding GCs, and explain these results as an interplay of radiative feedback and gas ejection during the star formation events. To compare the richness of GCs (i.e., the number of GCs per each galaxy) for different galaxy types, \cite{Zepf1993} introduced the quantity $T$ as the number of GCs per $10^9 M_\odot$ of galaxy stellar mass. Considering the metal-poor and the metal-rich populations separately, it would be possible to place more stringent constraints on the star-formation histories of galaxies \citep{KisslerPatig1997,Forbes2001}. In fact,  $T_{red}$ is expected to be significantly smaller for early elliptical galaxies compared to elliptical galaxies, given that they are expected to be formed through violent and gas-rich mergers with metal-rich GCs being formed within \citep{Brodie2006}. Instead, in the high-density environment, collapses are expected to form metal-poor GCs first, meaning that $T_{blue}$ is expected to be larger in hierarchical structure formation \citep{Rhode2005}. 

Modelling the evolution of GC populations requires, on one hand, the ability to describe how the galactic field affects the GC evolution in terms of tidal mass loss, shocks, dynamical friction and, on the other hand, the capability to closely follow GC internal dynamics, which regulates the GC mass loss, stellar population, compact remnants. Indeed, the GCs properties are outlined by  the  internal dynamical processes and by their host galaxy's evolutionary history \citep{Grudic2022,Rodriguez2022}. In this work, we present a novel approach that combines the MASinGa semi-analytic tool with the MOCCA-Survey Database I. MASinGa is a semi-analytic tool that performs population synthesis of GCs. These are evolved via a set of analytical fitting formulae describing the GC orbital evolution. MASinGa basics scheme was presented in \cite{ArcaSedda2014} and further improved in our companion paper \citep{ArcaSedda2022}. The MOCCA-Survey Database I contains realistic GC models performed with MOCCA, a Monte Carlo code that follows the long-term dynamical evolution of spherically symmetric stellar clusters, based on H\'enon's Monte Carlo method \citep[and references therein for details about MOCCA code]{Henon1971,Stodolkiewicz1982,Stodolkiewicz1986,Giersz2013}, together with stellar and binary evolution and strong interactions.
In this paper, we use MASinGa and MOCCA to create synthetic GC populations for the Milky Way (MW) and Andromeda (M31). We compare simulations with data from the Harris catalogue \citep[updated 2010]{Harris1996} and Baumgardt catalogue \citep{Baumgardt2018} for MW, and from the Revised Bologna Catalogue (RBC; \cite{Galletti2004,Galleti2006,Galleti2014}) for M31 respectively.

The paper is organised as follows: in Section \ref{sec:Method} we provide the details of the methodology and the physical recipes adopted in our tool, together with the extension introduced by this paper. In Section \ref{sec:InitCond} we describe the initial conditions used to reproduce the MW and M31 GC populations, with the obtained results presented in Section \ref{sec:Results}. Finally, in Section \ref{sec:Discussion} we provide our final conclusions and describe our future work. In Appendix \ref{Appendix} we show the results for M31's GC population derived from an analytical fit to the observed galaxy's rotation curve.

\section{Method} \label{sec:Method}
In this section we briefly discuss the main features of MASinGa and the MOCCA  Survey Database I.

The GC populations have been simulated using the MASinGa software \citep{ArcaSedda2022}. The  semi-analytic modelling  used in MASinGa has been used previously to carry out GC infall scenario studies \citep{ArcaSedda2014-2,ArcaSedda2015}. The GC infall scenario has been discussed as a process for the formation of a compact nucleus in the centre of a galaxy \citep{Tremaine1975,CapuzzoDolcetta1993,ArcaSedda2014-2}. Similarly, infalling GCs can merge in the galactic centre, enhancing the TDE event rate \citep{ArcaSedda2015} and contributing to the mass evolution of the SMBHs. 

\subsection{MASinGa code}
MASinGa (Modelling Astrophysical Systems In GAlaxies) enables the initialization of a GC system  for a given set of galaxy parameters. For each GC in the sample, MASinGa simulates the orbital evolution taking into account the galactic tidal field and shocks, which contribute to the cluster disintegration, dynamical friction, which drags the cluster toward the galactic centre, and internal relaxation, which regulates the cluster mass loss and expansion/contraction. MASinGa offers a wide series of choices for the galaxy parameters, the GC mass and spatial distribution, the total mass in GCs of a galaxy. An early version of MASinGa has been used to model the formation of NCSs via orbital segregation and merger of massive star clusters, a mechanism known as dry-merger.

In the following we describe the main features of MASinGa and the parameter chosen in this work. More details about the code are discussed in our companion paper \citep{ArcaSedda2022}.

\subsubsection{Galaxy density model and GC initial conditions}
In MASinGa, galaxies are modelled through the \cite{Dehnen1993} family of potential density pairs, characterised by a spherically symmetric density profile in the form:
\begin{equation*}
    \rho_G(r) = \frac{(3-\gamma)M_g}{4\pi r_g^3} \left( \frac{r}{r_g} \right)^{-\gamma} \left( 1 + \frac{r}{r_g}\right)^{\gamma-4},
\end{equation*}

where $M_g$ is the galaxy total mass in $M_\odot$, $r_g$ the galaxy length scale in kpc and $\gamma$ the density profile slope. Hereafter, we identify with $M_g(r)$ the galaxy mass enclosed within a galactocentric distance $r$. The advantage of the Dehnen density profile is the simple analytic form and the flexibility to generate different galaxy density profile distributions, determined by only two parameters $r_g$ and $\gamma$. The galaxy profile can be more or less cuspidal with the adjustment of those two parameters. 

MASinGa offers several choices to initialise galactic star cluster systems in terms of cluster mass function, radial distribution, or formation time. In our models, each GC is characterised by its galactocentric distance $R_{GC}$, its mass $M_{GC}$, the  eccentricity of the orbit $E_{GC}$, and the  half mass radius $r_{h,GC}$.

For our purposes, in this work we assume that the GC population is initially distributed across the galaxy following the density distribution of the host galaxy, with the GCs' total mass population being a fraction of the total galaxy mass, i.e. $\rho_{GCS}(r) = \alpha\cdot \rho_G(r)$. This implies that the total GC mass within any concentric radial annulus should be proportional to the total galaxy mass within the same radial annulus, with the total number of GCs within each radial bins set by the total GC mass inside the radial bin and the initial mass function (GCIMF). The GC masses $M_{GC}$ are selected randomly from the GCIMF. The orbital eccentricities $E_{GC}$ have been randomly picked from a thermal distribution. Finally, the $R_{GC}$ have been chosen randomly within the radial bins in which the galaxy density profile has been divided. This method ensures that in our initial model the density profile of the galaxy and the GC system share the same functional form.

\subsubsection{ Globular cluster dynamical evolution}
The interplay between the internal dynamics, the galactic tidal field, and the dynamical friction dictates the evolution and the survival of the GCs. Meanwhile the GCs internal evolution is driven by the stellar evolution and the relaxation process, the galactic tidal dissolution is driven by the change of the galactic gravitational field in which the GC moves. The presence of a tidal field influences the half-mass radius relaxation time-scale and the system mass loss, with a GC in a strong tidal field dissolving faster than an isolated GC. For a GC in a circular orbit and in a point-mass galaxy potential, the GC experiences a static tidal field. Internal evolution is dominated by stellar evolution, which regulates mass loss in the first $\sim 10-100$ Myr, while external evolution is regulated by the galaxy tidal field, which drives the cluster dissolution, and dynamical friction, which drags GCs toward the host galaxy centre \citep{Tremaine1975,CapuzzoDolcetta1993,Antonini2012,ArcaSedda2014-2}. The GC dissolution can also be boosted by  interactions with the galactic central regions and structural elements, like a bulge or stellar disk. These strong interactions are usually referred to as bulge and disc shocks. Based on the efficiency of the energy transferred during these phases, the dissolution can be catastrophic or diffusive, with the GCs disrupted in shorter or on longer time scales, respectively. If the galactic local density would be larger than the GCs' densities, the interaction between the GC and the galaxy could be catastrophic, resulting in the GC's dissolution.

As mentioned above, the dynamical friction drags the GCs towards the galactic centre, where they can contribute to the formation of a NSC \citep{Tremaine1975}. In previous studies, \citep{ArcaSedda2014,ArcaSedda2015} exploited N-body simulations of Dehnen galaxy models to derive a fitting formula for the dynamical friction timescale $\tau_{df}$ in the form:
\begin{multline}
    \tau_{df} = 0.3  \cdot g(E_{GC},\gamma)\cdot\left( \frac{r_g}{1\,kpc}\right)^{3/2} \left( \frac{M_g}{10^{11} M_\odot}\right)^{1/2} \cdot \\ \cdot  \left(\frac{M_{GC}}{M_{g}}\right)^{-0.67} \left(\frac{R_{GC}}{r_g}\right)^{1.76},
    \label{Eq:dynFriction}
\end{multline}
with $r_g$ in kpc and $M_{g}$ the total galaxy mass in $M_\odot$, $M_{GC}$ the GC mass  and  $R_{GC}$ its galactocentric position. The function $g(e,\gamma)$ is a dimensionless function given by
\begin{equation}
    g(e,\gamma) = (2-\gamma) \left[ a_1 \left(\frac{1}{(2-\gamma)^{a_2}} + a_3\right) (1-e) + e \right],
    \label{Eq:g_e_gamma}
\end{equation}
with $a_1=2.63\pm 0.17$, $a_2=2.26\pm0.08$ and $a_3 = 0.9\pm 0.1$ \citep{ArcaSedda2015}.

The importance of these three factors (internal dynamics, static and dynamical tides, and dynamical friction) is established by their respective timescales. The long term time-scale of internal dynamics is connected to the half-mass radius relaxation time-scale. If the dynamical friction time-scale is smaller than the GCs' age and the dissolution time-scale, the GCs would dissolve, possibly polluting the galactic halo. On the other hand, if the dissolution time-scale (connected with interplay between the relaxation process and tides) is smaller than the dynamical friction time-scale, the GCs would be dissolved before merging into galaxy center. In MASinGa, the star cluster orbital evolution is performed taking into account the orbital segregation driven by dynamical friction, the tidal dissolution driven by internal dynamics, the galactic tidal field, and the close orbital passages around the galactic bulge and across the galactic disc.  Moreover, to take into account the mass loss triggered by these disruptive mechanisms, in MASinGa we assume that the GC mass evolution follows an exponential form, $M_{GC}(t) \propto exp\,(-t/t_d)$ \citep{Henon1961}, where $t_d$ is the smallest disruption timescale among internal evolution, bulge and disc shocks, galactic field.

The galactocentric position time evolution is described by the dynamical friction time-scale evolution. In fact, the actual galactocentric position $r(t)$ at each time $t$ is given by 
\begin{equation*}
    \tau_{df} (r_0) - \tau_{df} (r) = t,
\end{equation*}
with $r_0$ being the initial galactocentric position and $\tau_{df}$ the dynamical friction time-scale, described by Eq. \ref{Eq:dynFriction} \citep{ArcaSedda2015}. Substituting the value for $\tau_{df}$, it is possible to determine the galactocentric position at time $t$. Finally, the eccentricity time evolution is described as $E_{GC}(t) = E_{GC}(t=0) \cdot  exp\,(-t/\alpha)$, with $\alpha = g(0,\gamma)/(g(E_{GC},\gamma)\cdot \tau_{df})$ as the orbit circularization time-scale due to dynamical friction, and $g(e,\gamma)$ as described in Eq. \ref{Eq:g_e_gamma}. This choice ensures that the orbit circularizes as the GC approaches the centre, as is expected from dynamical friction \citep{Colpi1999}.

In this work, we assume that all GC form at redshift $z= 4$, thus MASinGa evolves GC orbit either down to $z=0$, up to the cluster dissolution, or until the cluster orbit reaches the inner 10 pc, i.e. twice the observed half light radius of the MW NSC \citep{Chatzopoulos2015}. Also, models that were dissolved at distances within twice the chosen NSC radius (that is, within 20 pc) were considered as accreted to the NSC. Indeed, the dissolved GC would be gravitationally bound to the NSC and hence accreted. The possibility to follow GC dynamics down to the inner few pc of the galaxy enables us to place constraints on the possible formation of a NSC. To test the uncertainties in our models, we varied the maximum distance below which a GC is considered accreted into the NSC, finding not significant changes in the range $10-50$ pc. Finally, a GC was considered  disrupted if the local galactic density is greater than the GC half-mass radius density, or when the actual mass is smaller than 5\% its initial value.

\subsection{Updated GC internal dynamics recipes} \label{subSec:GCProp}
Semi-analytic tools like MASinGa offers the unique advantage of a risible computational load, thus allowing the realisation of hundreds galaxy models within a few hours. Nonetheless, the simplistic approach behind this type of tools misses the great level of detail attainable with direct N-body and Monte Carlo codes, whose computational costs made, however, impossible any population studies. MASinGa is conceived and devised to efficiently exploit the advantages of both semi-analytic and N-body methods. In fact, MASinGa can be interfaced with cluster simulation catalogues to provide a comprehensive view of internal and external GC evolution. In this work, we devise an interface to couple MASinGa with the MOCCA-Survey Database I, as explained in the following.

The Monte Carlo methods are known to be fast and reliable, with results comparable with the NBODY ones \citep{Wang2016,Kamlah2021} and MW GC properties \citep{Leveque2021,Leveque2022}. The models from the MOCCA-Survey Database I \citep{Askar2017}  have been used in this work. In order to couple the MASinGa results with the MOCCA-Survey Database I model, an update for the internal dynamical evolution has been applied to the MASinGa analytical equations. Better estimations of mass loss, tidal field, and half-mass radius have been introduced in this work, together with the creation of the MASinGa-MOCCA model connection.

The tidal radius and half-mass radius evolution play an important role in the GC mass loss. More compact clusters (that is, with small half-mass radius versus tidal radius ratio) would remove less objects (stars or binaries) from the system, and therefore loose less mass compared to less compact clusters. MOCCA-Survey Database I models take into account the realistic evolution of star clusters and do show a different mass loss evolution for tidally filling and underfilling models. To better represent the mass loss for MASinGa models at 12 Gyr, and to compare GC models from MASinGa with the ones from MOCCA Database I, a better constraint for mass loss (and hence half-mass radius and tidal radius) evolution was needed.

In MASinGa,  the initial tidal radius $r_{tidal}$ is determined from the galactocentric distance $R_C$. Supposing a circular orbit, the given $r_{tidal}$ in the external galaxy is calculated according to the equation:
\begin{equation} \label{eq:Rt}
R_C: \qquad r_{tidal} = R_C \cdot \sqrt[3]{\frac{M_{GC}}{3 M_{g}(R_C)} },
\end{equation}
with $M_{g}(R)$ as the galaxy's mass at position $R$. The galaxy's mass at each position can be determined by the galaxy's rotational curve or from the Dehnen model mass profile.

The initial half-mass radius for MASinGa models are determined by the \cite{Marks2012} relationship. In order to reproduce the observed scatter, the initial half-mass radius value has been increased by a multiplicative factor, chosen randomly from a uniform distribution between 1 and 15. A minimum and a maximum value for $r_{h,GC}$  of 0.2 and 7 pc was set, respectively. An additional limitation has been imposed on $r_{h,GC}$, so that its value cannot be larger than 0.3 $r_{tidal}$. In fact, for a cluster with an initial spatial distribution described by a tidally underfilling King model \citep{King1966} with $W_0=6.0$, the half-mass radius is around 10 times smaller than the initial $r_{tidal}$.

The evolution of the tidal radius at each time $t_i$ takes into consideration the mass loss of the system,
\begin{equation*}
    r_{tidal}(t_i) = r_{tidal}(t_0) \cdot \sqrt[3]{M_{GC}(t_{i-1})/M_{GC}(t_0)},
\end{equation*} with $M_{GC}(t_0)$ being the initial GC mass, $t_0 = 0$ and $t_i > t_{i-1}$. Instead, the evolution of the half mass radius followed Eq. 1 from \cite{Giersz1996}, 
\begin{equation*}
    r_{h,GC}(t_i) = r_{h,GC}(t_0)\cdot b\cdot(t_i - T_0)^{(2+\nu)/3}
\end{equation*} 
The value for $b$, $T_0$ and $\nu$ have been obtained by fitting the formula to the half-mass radius evolution to all the models used in Paper I \citep{Leveque2021}. The mean values are: $b = 0.787\pm 0.124 \,\,pc/Myr$, $\nu = -1.467 \pm 0.545$, $T_0 = -13.48\pm4.56$ Myr. The models used for our estimation are only models that survived 12 Gyr of evolution. This means that the reported values are biased for the surviving clusters only.

Finally, the mass evolution is determined by the Spitzer formula \citep{Spitzer1987}, so that
\begin{equation*}
    M_{GC}(t_i) = M_{GC}(t_0) \cdot e^{-t_{c}/t_i},
\end{equation*}
with $t_c = t_{relax}/\xi$ and $t_{relax}$ is the initial Spitzer relaxation time \citep{Spitzer1987}. The value of $\xi$ is determined by whether the model is isolated or not at each time step t, that is:
\begin{equation*}
\begin{alignedat}{2}
    \xi &= 8.5 \times 10^{-3}  \qquad & if \,\,r_{h,GC}(t_i)/r_{tidal}(t_i) & \leq 0.1\\
    \xi &= 4.5 \times 10^{-2}  \qquad & if \,\,r_{h,GC}(t_i)/r_{tidal}(t_i) & > 0.1
\end{alignedat}
\end{equation*}
For an initially tidally filling King cluster \citep{King1966} with $W_0=6.0$, the initial half-mass radius is 10 times smaller than the initial tidal radius. This value has been chosen as a limiting value to describe an isolated and non-isolated cluster in our simulation.  The MOCCA-Survey Database I models used in this study were mostly tidally-underfilling models, with  $r_{h,GC}(0)/r_{tidal}(0)= $ 0.02 and 0.04. So it seems that the mass loss from the cluster can be described, for a considerable fraction of its evolution, as for an isolated cluster. The assumption about mass loss according to Spitzer’s recipe for isolated clusters is a reasonable first-order approximation. However, due to the mass loss connected with stellar evolution and binary energy generation, the cluster will expand and eventually  will become tidally filling (after a relatively long time, depending on the  degree of underfilling). In that case, the mass loss is  governed by the galactic tidal field. This rather rough treatment should provide an approximate evolution of the mass loss from the cluster and also provide the evolution of the tidal radius.

\subsection{MASinGa-MOCCA connection} \label{subSec:MOCCA_connection}
At the end of the MASinGa evolution, the MASinGa results are coupled with the models from the MOCCA-Survey Database I. To better reproduce the observed properties of the studied GC populations, a model subset from the MOCCA-Survey Database I was chosen, with the selection procedure described in Sec. \ref{subSec:MOCCADatabase}. Using the subset of models from MOCCA-Survey Database I, a library of MOCCA models has been generated - the MOCCA-Library models. The MOCCA-Library consists of MOCCA-Survey Database I model representations with different orbital properties and orbital positions in the studied galaxy's gravitational field. Models from the MOCCA-Library were picked to populate the studied galaxies, and used to provide detailed GC observational properties.

\subsubsection{MOCCA models in a different tidal field}
In order to determine the galactocentric position in the studied galaxy gravitational potential, a few steps have been taken. First of all, the tidal radius and galactocentric distance for a circular orbit in the studied galaxy potential has been determined. The MOCCA-Survey Database I models were assumed to move on a circular orbit at Galactocentric distances between $1$ and $50$ kpc in the Galaxy. The Galactic potential was modelled as a simple point-mass, taking as the central mass the value of the galaxy mass enclosed within the GC's orbital radius. The GC's rotation velocity was set to $220$ $km\,\,s^{-1}$ over the whole range of galactocentric distances. Knowing the tidal radius for the MOCCA model and the density/potential distribution for the simulated galaxy, it is possible from Eq. \ref{eq:Rt} to determine the correct galactocentric distance $R_C$ for a circular orbit in an external galaxy, for a given tidal radius $r_{tidal}$ and GCs mass $M_{GC}$.

\cite{Cai2016} exploited N-Body simulations to establish the evolution of GCs on circular and eccentric orbits. The authors established the apocentric distance for the eccentric orbit that has a lifetime similar to a cluster with the same mass on a circular orbit. Therefore, for each GC eccentric orbit, it is always possible to find a circular orbit on which the GC will experience an equivalent mass loss. Fitting the data shown in their Fig. 6,  it is possible to find the apocenter distance  $R_{apo}$, scaled to $R_C$, for an eccentric orbit with an initial orbital eccentricity $E_{GC}$ as:
\begin{equation}
\frac{R_{apo}}{R_C} = (1 - 0.71 \cdot E_{GC})^{-5/3}. \label{Eq:Cai}
\end{equation}
The pericenter distance is then determined as $R_{peri} = 2.0\cdot a - R_{apo}$, with $a$ as the semi-major axis of the orbit. Using Kepler's 3rd law, we found the GC orbital period P, and then computed the mean anomaly $\epsilon$,
\begin{equation*}
    \epsilon = \frac{2 \pi \cdot (t - T)}{P},
\end{equation*}
with T being the periastron passage time and t the current time. The term $(t-T)$ has been randomly picked between 0 and P. Successively, the eccentric anomaly $E$ has been determined solving the Kepler equation, $\epsilon = E - E_{GC}\cdot sin(E)$. Knowing the relation between the eccentricity and the true anomalies $\mu$, 
\begin{equation*}
    \tan\left(\frac{\mu}{2}\right) = \sqrt{\frac{1+e}{1-e}} \tan\left(\frac{E}{2}\right),
\end{equation*}
the galactocentric position $R_{GC}$ has been determined as
\begin{equation*}
    R_{GC} = \frac{a \cdot (1-e^2)}{1+e\cdot cos(\mu)}
\end{equation*}
This simplistic procedure does provide a reasonable first order approximation for the GC's galactocentric distance distribution. As for the observed GC populations in external galaxies, the actual orbit and actual galactocentric distances are unknown. For this reason, the GC's galactocentric distance $R_{GC}$ has been chosen randomly within the orbit apocentre and pericentre.
    
\subsubsection{MOCCA-Library model generation and selection}
For each model in the selected subset, 30 eccentricities have been selected. Each eccentricity identifies an orbital representation in the studied galaxy (see Eq. \ref{Eq:Cai}). The MOCCA-Library eccentricity $E_{GC}$ has been randomly chosen from the thermal distribution. For each orbital representation, $R_{GC}$ has been sampled 30 times within the orbit apsis as described above\footnote{In a proper GC orbit representation, the orbit should be sampled with higher probability close to the pericenter and apocenter where the radial velocity is smallest.}, giving a total of 900 representations of each unique MOCCA-Survey Database I model in the MOCCA-Library. This procedure allows us to populate the same model in different regions of the galactic field (that is, different galactocentric radial bins), and with different orbital eccentricities. 

For a fixed galaxy model, the dynamical friction time-scale and the galaxy-GC interactions are functions of position and eccentricity only, as shown in Eq. \ref{Eq:dynFriction}. Therefore, the family of different representations in the studied galaxy implies a variety of dynamical interactions with the galactic potential field for each MOCCA-Survey Database I model. These model representations have similar mass loss as simulated in the MOCCA-Survey Database I model. In fact, a shorter dynamical friction time-scale would grab the GCs closer to the galactic center, and eventually disrupt them due to the high galactic density or even lead to accretion into the galactic center. Finally, models from the MOCCA-Library are defined as unique when they represent different MOCCA models. Indeed, for different unique models the internal dynamical evolution is diverse - for example, mass loss, half-mass radius, and compact-object composition. The different representations of one unique model in the MOCCA-Library result in a similar internal dynamical evolution, but with an important diversity in external dynamical evolution.

MOCCA models which lasted up to 12 Gyr were used in our simulations. In order to preserve the initial number of GCs and the number of surviving GCs, the number of GCs for the MOCCA populations are set equal to the number of MASinGa models that at 12 Gyr survived the internal dynamical evolution, independent of whether they were disrupted because of external interactions with the galaxy or sunk to the galactic center.

This means that each MASinGa model has to be mapped to a MOCCA-Library model representation, taking into account its initial position and initial mass. To simplify the model selection procedure and reduce the number of repetitions of unique MOCCA Database models, the MOCCA-Library population has been divided into a 2D matrix grid, according to the GCs' initial galactocentric distance and initial mass bins. For each MASinGa model, only one representation for each unique MOCCA model was randomly selected within each matrix cell: the same initial galactocentric bin from the MASinGa model has been used to select the matrix cell, meanwhile the initial mass bin was selected randomly from the GCIMF cumulative distribution. Finally, one model representation from the selected model representations has been randomly chosen to represent the considered MASinGa model and successively removed from the MOCCA-Library\footnote{Being $N_1$, $N_2$, etc., the total number of MOCCA-Library model representations in each cell and $N'_1$, $N'_2$, etc., the number of representations of unique models in each cell, only one model has been chosen among the representation ones $N'_1$, $N'_2$, etc. Whenever a representation model has been chosen to represent a MASinGa model, it has been removed from the MOCCA-Library. This means that the new number of model representations and unique models in the corresponding cell would be $N_i-1$ and $N'_i -1$, respectively.}. It is expected that multiple instances (even thousands) of MOCCA representations populate each mass and galactocentric position bin, implying multiple repetitions of the same unique MOCCA models. This procedure has been necessary to reduce the bias to over-reproduce the MASinGa models with few unique MOCCA models.

The GCIMF has been divided into only a few mass groups with different bin sizes, in order to contain at least one model from the MOCCA-Survey Database I. Indeed, as it will be explained in Sec. \ref{subSec:MOCCADatabase}, the initial masses in the MOCCA-Survey Database I are not continuous. Instead, the GCIMF cumulative distribution has been normalized to the initial MASinGa models mass distribution. This whole procedure will assure that the galaxy density distribution together with the GCIMF distribution  will reproduce the initial conditions in the MASinGa population. 

\begin{figure}
    \centering
    \begin{subfigure}{0.5\textwidth}
       \centering
        \includegraphics[width=\linewidth]{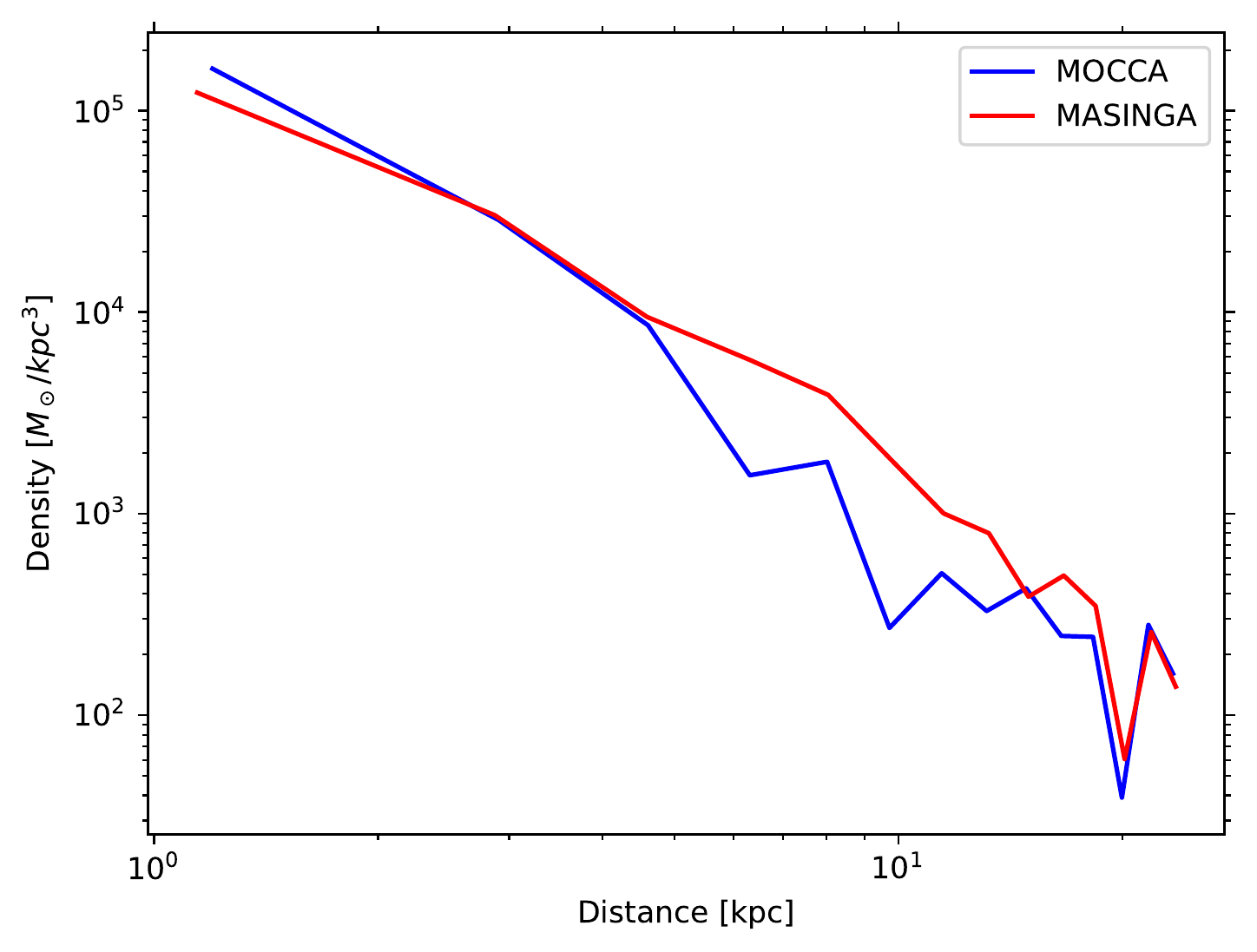}
    \end{subfigure}
    \begin{subfigure}{0.5\textwidth}
       \centering
        \includegraphics[width=\linewidth]{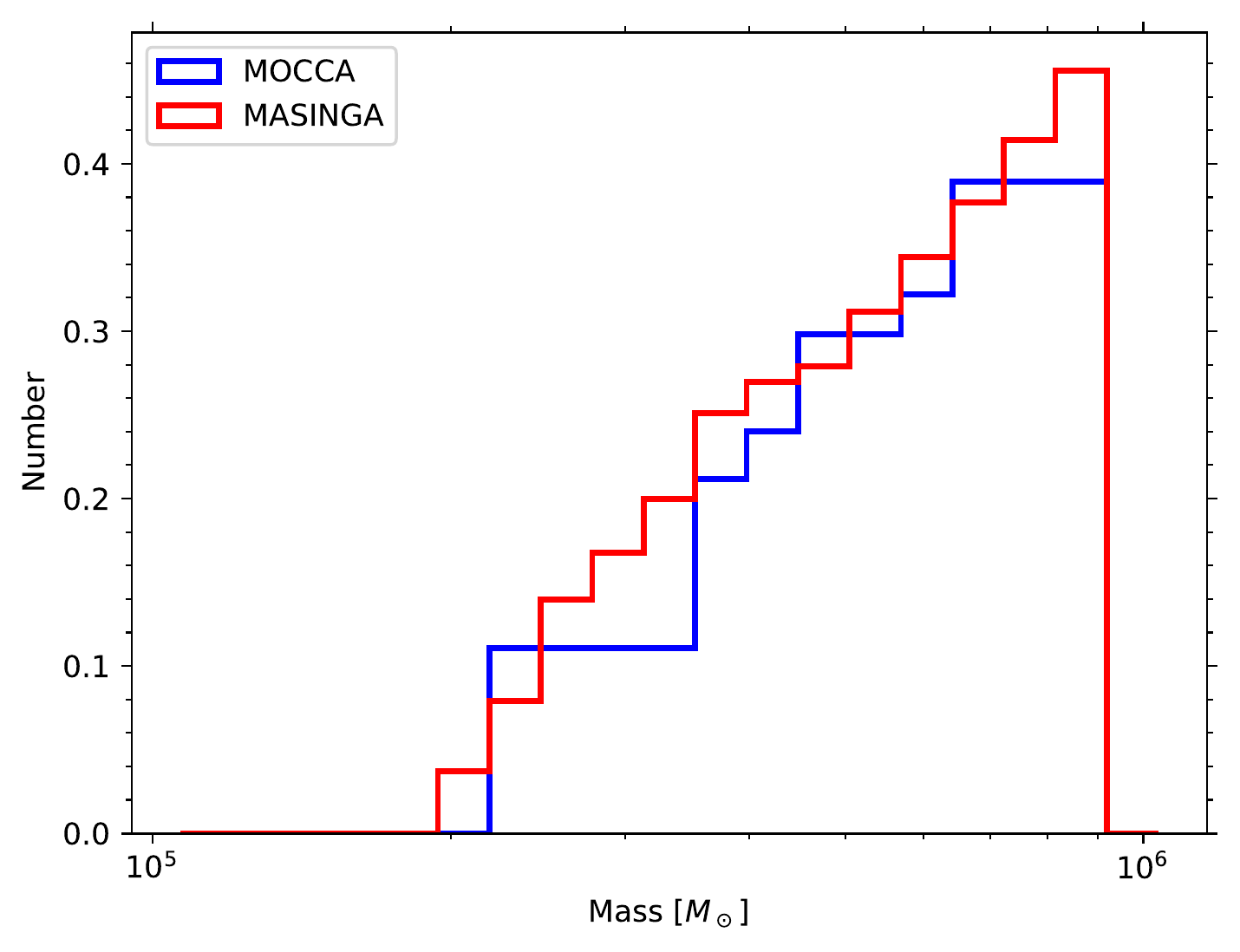}
    \end{subfigure}
    \caption{Initial density distribution (top) and GCIMF  cumulative distribution (bottom) for a randomly selected model for MOCCA (blue) and MASinGa (red) for the MW representation. Similar results have been obtained for the M31 representation.}
    \label{Fig:InitDensity}
\end{figure}

In Fig. \ref{Fig:InitDensity} we report the comparison between the MASinGa and MOCCA initial GCs density distribution and their GCIMF cumulative distribution. The GC density distribution in MASinGa follows the galaxy density distribution function, and the MOCCA-Library density distribution is comparable with the MASinGa one. Similarly, the GCIMF cumulative distribution inferred from the MOCCA-Library is in good agreement with the MASinGa distribution, taking into consideration the discontinuity in the initial mass in the MOCCA-Survey Database I models. 

\subsubsection{MOCCA-Library model evolution}
The connection with the MASinGa code is necessary to describe the GCs' dynamical friction and the interactions with the galaxy evolution with the internal properties for the MOCCA models being already known. Once the MOCCA-Library models have been connected with the MASinGa GC population, the selected models were evolved. The actual mass, half-mass radius, and  internal objects' property values from the MOCCA-Survey Database I were used to determine their evolution within MASinGa. Instead, the galactocentric distance and eccentricity evolution are determined using the equations  adopted in MASinGa, as described in Sec. \ref{subSec:GCProp}. Finally, the same condition for sinking to the NSC used for MASinGa models were applied to the chosen MOCCA-Library models - that is models have been considered accreted to the NSC if their galactocentric position is smaller than the NSC radius. Also, the models were considered as disrupted if the local galactic density was found to be greater than the GC half-mass radius density.

\section{Initial conditions} \label{sec:InitCond}
\subsection{Observed GC populations for MW and M31}
The results from our simulations have been compared with the observational data from the Harris catalogue \citep[updated 2010]{Harris1996} and Baumgardt catalogue \citep{Baumgardt2018} for the MW, and from Revised Bologna Catalogue (RBC; \cite{Galletti2004,Galleti2006,Galleti2014}) for M31.

The Harris catalogue contains the basic parameters for the 157 classified GCs observed in the MW galaxy. Meanwhile, the structural parameters for the 112 MW GCs are reported in the Baumgardt catalogue. The RBC  contains the 231 confirmed M31 GCs and their positions, photometry, velocities, structural parameters, metallicities, and lick indexes. The complete catalogue also contains information about non-GC objects (such as galaxies or GC candidates). In this work, we selected only objects in the catalogue that are confirmed GCs.

Additionally, for the M31 GC population, the observed $V$ magnitude has been transformed to the absolute $V_{abs}$, using a distance of M31 from the Sun of $783.43$ kpc and $E(V-B) = 0.11$, as reported in \cite{Galletti2004,Galleti2006}.The galactocentric distances have been determined as the distance between the position in the sky of each GC and the M31 centre ($RAJ2000 = 00 \,42\, 44.330,\,\, DECJ2000 = +41 \,16 \,07.50$).

The total mass for each GC in the Harris catalogue and in the RBC was estimated using a mass-to-light ratio $M/L_V = 1.83$ \citep{Baumgardt2020}, with $L_V$ as the absolute V luminosity expressed in units of $L_\odot$ and M in units of $M_\odot$. The dispersion around the $M/L_V$ mean found in   \cite{Baumgardt2020} of $0.24$ was used to determine the mass error for each GC. 
The determined GC masses range between $1\times10^4 - 2\times10^6$ and $5\times10^4 - 3\times10^6$, for MW and M31 respectively.

As seen in most of the observed galaxies, the largest parts of the GC populations are located within a few kpc from the galactocentric centre. Furthermore, due to observational limits and errors, detecting and confirming GCs that would belong to external galaxies' GC populations at larger distances from the galactic centre can be challenging. Indeed, the number of confirmed GCs in the RBC catalogue are distributed within 17 kpc from the galactic centre. For this reason, we limited our study to GCs within 17 kpc from the galactic centre. 

The derived structural parameters (such as half-light radius, core radius, etc.) for M31 have been derived by fitting to the surface brightness profile of the observed GCs. The derived parameter uncertainties are enhanced for smaller GC surface brightness profiles. In order to reduce the uncertainties of the half-light radius, we have considered only the observed M31 GCs with half-light radius surface brightnesses (defined as $L_V / r_h^2$, with $L_V$ being the total $V$ luminosity and $r_h$ the half-light radius) greater than $4000\,\, L_\odot/pc^2$. This value was set arbitrarily, but with the aim of keeping  a large fraction of the observed GCs. Indeed, $\sim 95\%$ of the M31 GC population has a half-light radius surface brightness greater than this value.

\subsection{MOCCA-Survey Database I model selection} \label{subSec:MOCCADatabase}
The MOCCA-Survey Database \citep{Askar2017} consists of nearly 2000 real star cluster models that span a wide range of initial conditions, provided in Table 1 in \cite{Askar2017}. For half of the simulated models, supernovae (SNe) natal kick velocities for neutron stars (NSs) and BHs are assigned according to a Maxwellian distribution, with a velocity dispersion of $265 \,\,km \,s^{-1}$ \citep{Hobbs2005}. In the remaining cases, BH natal kicks were modified according to the mass fallback procedure described by \cite{Belczynski2002}. Metallicities of the models were selected as follows: $Z = 0.0002,\, 0.001,\, 0.005,\, 0.006, \,0.02$. All models were characterized by a \cite{Kroupa2001} IMF, with minimum and maximum initial stellar masses of $0.08$ and $150 \,M_\odot$, respectively. The GC models were described by the \cite{King1966} profile with central concentration parameter values $W_0 = 3, 6, 9$. They had tidal radii ($r_{tidal}$) equal to: 30, 60, or 120 pc, and were either tidally filling or had ratios between $r_{tidal}$ and the half-mass radius ($r_h$) equal to 25 or 50. The primordial binary fractions were chosen to be $5\%,\, 10\%,\, 30\%$, or $95\%$. Models characterized by an initial binary fraction equal to or lower than 30\% had their initial binary eccentricities selected according to a thermal distribution \cite{Jeans1919}, with mass ratios and  logarithms of the semi-major axes according to uniform distributions. For models containing a larger binary fraction, the initial binary properties were instead selected according to the distribution described by \cite{Kroupa1995}, via so-called eigen-evolution and mass feeding algorithms. The models consist of $4 \times10^4,\, 1 \times10^5,\, 4 \times10^5, \,7 \times10^5,\,1.2 \times10^6$ objects (stars and binaries). As shown by \cite{Askar2017} and \cite{Leveque2021}, the MOCCA models reproduce observational properties of Milky Way GCs relatively well.

\subsection{MASinGa initial conditions} \label{subSec:initCond}

Dynamical friction is an important process for the GC galactocentric distance evolution. As shown in \cite{ArcaSedda2014} and Eq. \ref{Eq:dynFriction}, it strongly depends on the galaxy mass $M_g$, typical radius $r_g$, and slope of the matter density $\gamma$. In Dehnen models \citep{Dehnen1993}, these quantities can be used to calculate the rotation curve, given in the form:
\begin{equation*}
    v_c^2(R) = G \cdot M_g \cdot \frac{R^{2-\gamma}}{(R + r_g)^{3-\gamma}},
\end{equation*}
with $R$ being the galactocentric distance and $G$ the gravitational constant.

\begin{figure}
    \centering
    \begin{subfigure}{0.5\textwidth}
       \centering
        \includegraphics[width=\linewidth]{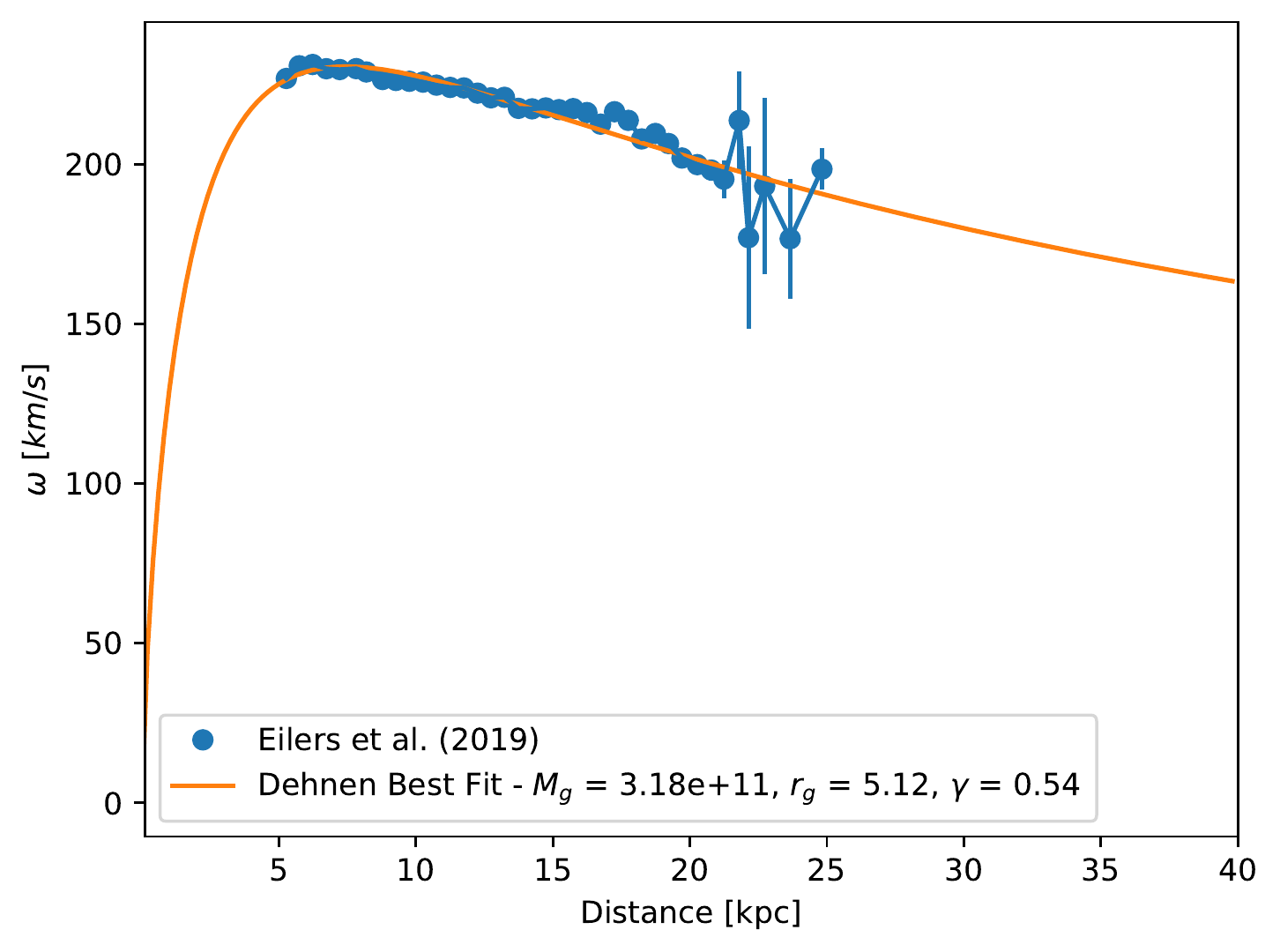}
    \end{subfigure}
    \begin{subfigure}{0.5\textwidth}
       \centering
        \includegraphics[width=\linewidth]{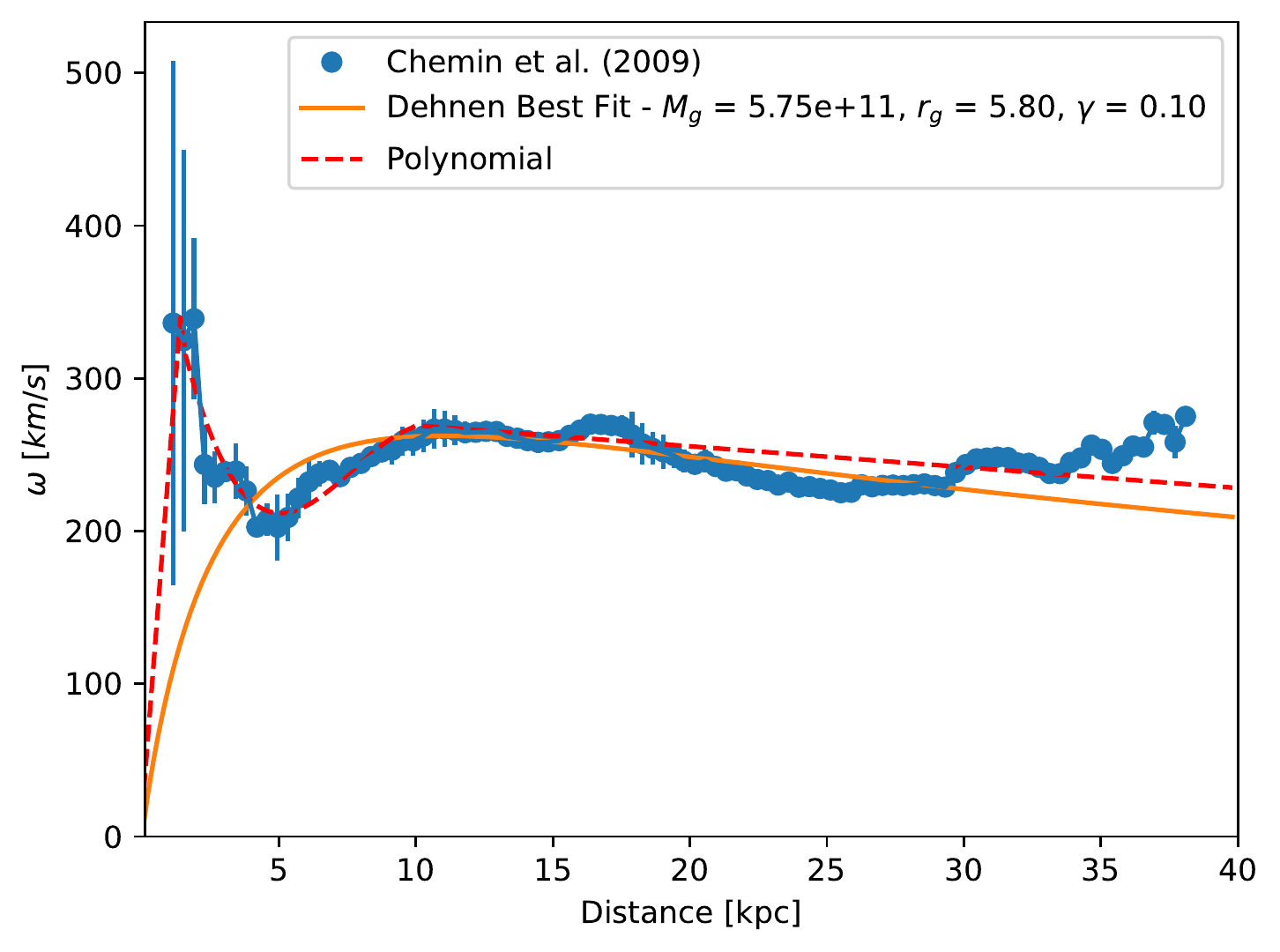}
    \end{subfigure}
    \caption{Rotational curve fitting for MW (top) and M31 (bottom). In the figure label, the Dehnen best fit parameters are reported.}
    \label{Fig:rotCurve}
\end{figure}

We fit the equation above to tailor our model to the MW and M31 rotation curves, finding the best fit parameters to the observed MW rotational curve \citep{Eilers2019} are $M_g = 3.18\times 10^{11}\,\,M_\odot$, $r_g = 5.12$ kpc, $\gamma = 0.54$.  Similarly, the best fit to the observed M31 rotational curve \citep{Chemin2009} gives $M_g = 5.75\times 10^{11}\,\,M_\odot$, $r_g = 5.8$ kpc, $\gamma = 0.1$. The best fit, together with the observational rotational curves are showed in Fig. \ref{Fig:rotCurve} with the top and bottom panels corresponding to the MW and M31, respectively. The Dehnen density profile cannot reproduce the observed increase of the observed rotational velocity in the central region of M31 ($\sim 5$ kpc), implying a mass underestimation in that zone of the galaxy and, as will be also discussed in Appendix \ref{Appendix}, an underestimation in the number of infalling GCs in the central region, the NSC mass growth, and the mass evolution of the infalling IMBHs in the NSC. To better understand the importance of the underestimation of the GCs infall, a polynomial fit was applied to the M31 rotational curve to reproduce the central density increase in the observed rotational curve. In particular, the polynomial curve was divided into three regions to better reproduce the observed curve: a linear fit between $0$ and $1.5$ kpc, a cubic fit between $1.5$, and $10$ kpc and a linear fit between $10$ and $20$ kpc. In this model, the polynomial curve fit has been used to determine the mass and the density profile of the simulated M31 galaxy. Instead, the Dehnen best fit model parameters for M31 rotational curve have been used to determine the dynamical friction time-scales, as described in Eq. \ref{Eq:dynFriction}.

The initial mass function for GCs is set to be a powerlaw $dN/dm = b \cdot m^{-\alpha}$ with a slope of $\alpha = 2$ \citep{Lada1991,Kroupa2001}. The GC's total mass population is expected to be a fraction of the total galaxy mass, that is $M_{GCS} = \beta M_g$. The $\beta$ parameter can be estimated from the observed GC masses. \cite{Webb2015} found that the initial GC mass was $\sim 5$ times larger than the actual observed values. This results is in agreement with the values reported in the MOCCA Database I, with a mean mass ratio at the initial time and at 12 Gyr being 4.2. Using this result, we determined the initial minimum and maximum masses for the observed GCs in MW and M31, being $M_{min} = 4.2 \times 5 \times 10^4 \sim 2.1 \times 10^5 M_\odot$ and $M_{max} = 4.2\times3\times10^6 \sim 1\times10^7 M_\odot$, respectively. The considered observed GC populations (located within 17 kpc from the galactic centre) have total masses of $M_{GCS,17} \sim 3\times 10^7\,\,M_\odot$ and $ \sim 1\times 10^8\,\,M_\odot$, for MW and M31, respectively. The total galactic mass included within 17 kpc is obtained from the best fit to the rotational curve, giving $M_{g,17} \sim 10^{11}\,\,M_\odot$ and $ \sim 2.5 \times 10^{11}\,\,M_\odot$, for MW and M31, respectively. For the observed GC populations in MW and M31 we obtained a value of $\beta = M_{GCS,17} / M_{g,17} \sim 10^{-3}$ for both MW and M31. 

However, the $\beta$ parameter used in our simulations has to be adjusted. In fact, the maximum MOCCA initial mass is $1.1\times10^6 \,\,M_\odot$, much smaller than the initial mass seen in the observations ($M_{max} = 1\times10^7 M_\odot$). As expressed previously, the minimum mass is set to reproduce a minimum mass at 12 Gyr of $5\times10^4$, implying an initial mass of $2\times10^5$. The total GC mass simulated from the MOCCA models would be
\begin{equation*}
    b \cdot \int^{1.1\times10^6 M_\odot}_{2\times10^5 M_\odot} m^{1-\alpha} dm = M_{GC,MOCCA} = \beta_{MOCCA} M_g
\end{equation*}

To properly scale the total mass population, an appropriate value of $\beta_{MOCCA}$ has been calculated as
\begin{equation}
\begin{alignedat}{2}
    \beta_{MOCCA} = \beta \cdot \frac{\int^{1.1\times10^6 M_\odot}_{2\times10^5 M_\odot} m^{1-\alpha} dm }{\int^{M_{max}}_{M_{min}} m^{1-\alpha } dm}.
\end{alignedat}
\end{equation}
A value of $\sim10^{-4}$ is obtained, and specifically a value of $2.0\times10^{-4}$ and $3.0\times10^{-4}$ has been used to determine the total GC population masses within 17 kpc during the MASinGa initial conditions for MW and M31, respectively.

\section{Results} \label{sec:Results}
To filter out statistical fluctuations we create 100 galaxy models for MW and M31, all formed 12 Gyr ago and evolved until present day as described above. We considered the mean values obtained from those GC populations to better obtain a more robust statistical representation of the models. 

\subsection{Comparison with observations}
In this paper we introduced the machinery that will be used in the following works to populate in an automatic way the GC populations of thousands of galaxies in the local universe. In order to reproduce the observed properties of the external galaxies, limitations in the simulated models (such as masses and galactocentric distances) are applied. In particular, in this work we restrict the selected models from the MOCCA-Survey Database I to generate the MOCCA-Library used in MASinGa to those in which the fallback prescription \citep{Belczynski2002} was used. In order to generate the initial conditions to be as generic as possible and following the observational mass at 12 Gyr, models with masses at 12 Gyr larger than $5\times10^4 M_\odot$ have been selected to reproduce both MW and M31 GC populations. Similarly, 
we have considered MOCCA models with half-light radius surface brightnesses greater than $4000\,\, L_\odot/pc^2$.

As output from the MASinGa code, the 3D galactocentric distance for each GC is given. On the other hand, the observed distances for MW and M31 are projected distance in the sky plane. To compare our models and observations, we project MASinGa cluster position onto the plane of the sky.

\begin{figure*}
    \centering
        \includegraphics[width=\linewidth]{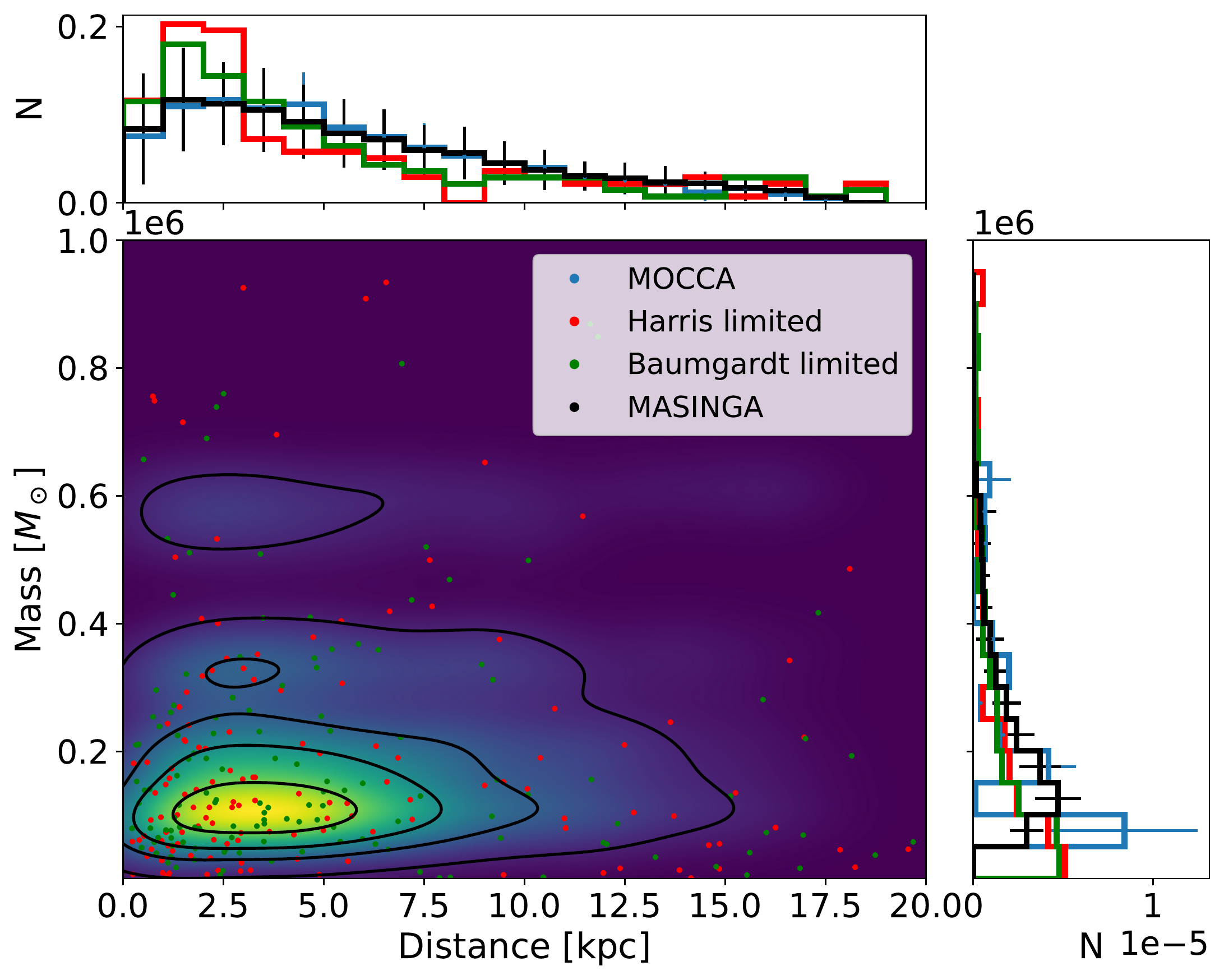}
    \caption{Density map for the galactocentric position and final mass for the MOCCA population for MW. The contours include the 80, 50, 30 and 10\% levels of population. The Harris and Baumgardt catalogues are reported in red and in green, respectively. On the side, the normalized histogram showing the distributions of each population is reported, with the error bars showing the standard deviations for the simulated models. In blue and black lines, the histogram for MOCCA and MASinGa models are shown, respectively.}
    \label{Fig:densityMapMW}
\end{figure*}

\begin{figure*}
    \centering
        \includegraphics[width=\linewidth]{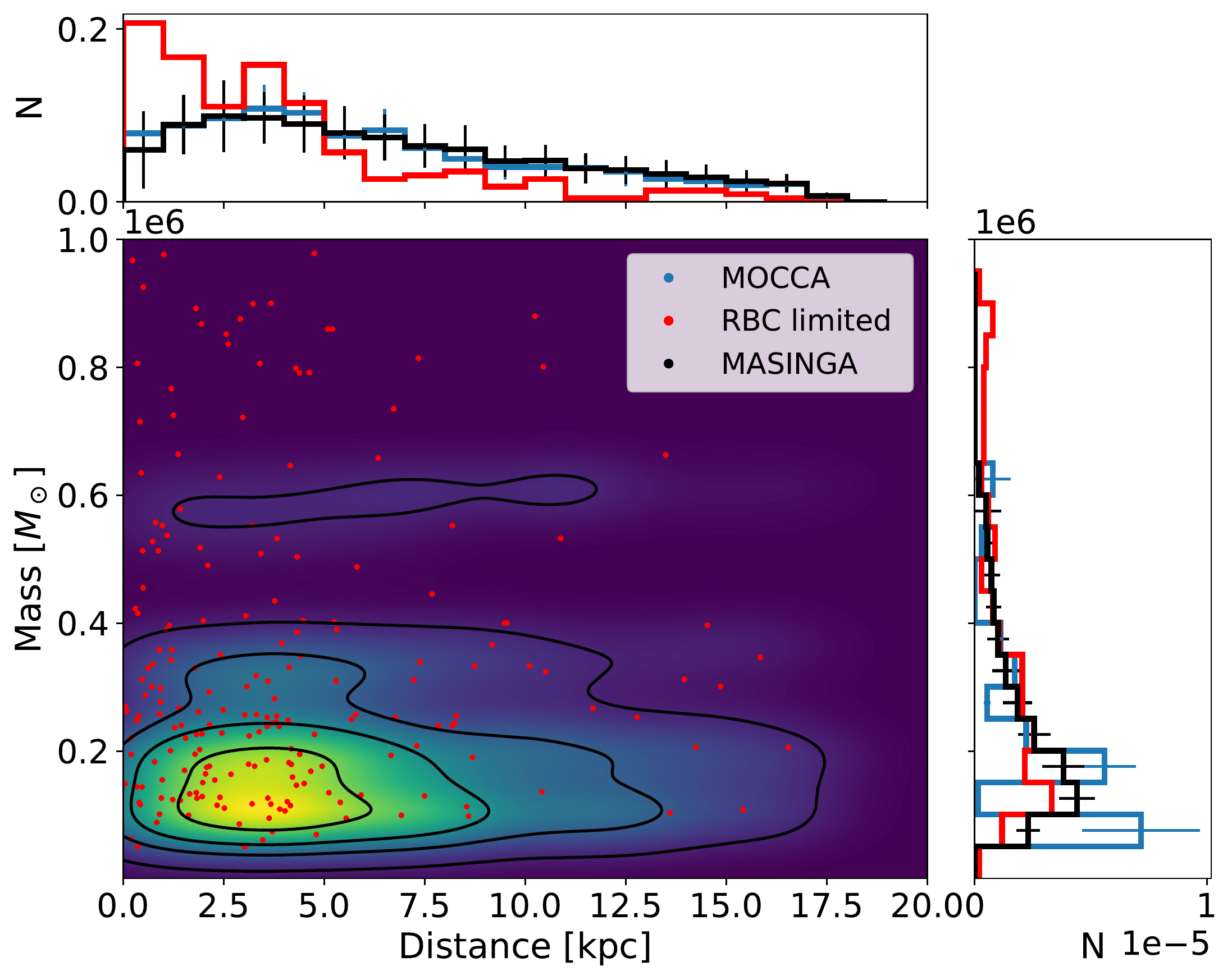}
    \caption{Density map for the galactocentric position and final mass for the MOCCA population for M31.  The contours include the 80, 50, 30 and 10\% levels of population, respectively. The RBC catalogue is reported in red. On the side, the normalized histogram showing the distributions of each population is reported, with the error bars showing the standard deviations for the simulated models. In blue and black lines, the histogram for MOCCA and MASinGa models are shown, respectively.}
    \label{Fig:densityMapM31}
\end{figure*}

In Fig. \ref{Fig:densityMapMW} and \ref{Fig:densityMapM31}  the density map for the final galactocentric position and final mass for the MOCCA population for MW and M31 respectively has been reported. The colour map shows the density map for our models, and the contours include the 80, 50, 30 and 10\% levels of population. In black we reported the results from MASinGa, while in red, we reported the properties of MW and M31 GCs retrieved from the Harris and RBC catalogue, respectively. For MW, the observed properties from the Baumgardt catalogue are reported in green. The regions containing most of the populations are presented with brighter color. The comparison shows that our models represent decently well the galactocentic distance and mass distribution of MW and M31 clusters. As expected, most of the GC populations are localized at smaller galactocentric distances, with a decreasing number of GCs at larger galactocentric positions. The overall spatial distributions obtained from our models are in  relatively good agreement with the observed ones. However, our models exhibit a lower number of clusters within 5 kpc compared to observations. As it will be discussed in Sec.  \ref{sec:Discussion}, these differences can be due to the bimodal nature of the observed GC populations, which has not been simulated in our models. Finally, both observed and simulated GCs have mostly masses in the range $10^5$ and $3\times10^5 M_\odot$, with a small percentage having large masses ($> 5\times10^5 M_\odot$). Our simulated populations show a mass distribution in relatively good agreement with the observed ones. The peaks seen in the simulated distributions are connected to the mass distribution in the MOCCA models.

In \cite{Madrid2017}, the authors studied the mass loss and evaporation rate of GCs in a strong Galactic tidal field, as a function of time and galactocentric distance. Their N-body simulations were compared to the MOCCA models. The authors found that the mass loss in the inner Galactic region can be enhanced. The MOCCA models were comparable with N-body simulation evolution for Galactocentric distances down to few kpc. Given that, the MOCCA models have been simulated for a constant rotational velocity and from the observed rotational curve, it is possible to see an almost constant rotational velocity in the galacatocentric distance between $\sim 5$ and $20$ kpc. Because of the point-mass approximation used in the simulations, it is expected that MOCCA models would not reproduce the mass loss of GCs at smaller distances. In order to not limit too strongly the comparison sample, the post-processing investigation and the statistical determination of GC populations' properties have been carried out for the region between $2$ and $17$ kpc.

One possible source of bias in our models owes to the fact that the number of MOCCA models is finite, thus it can happen that a MOCCA GC is used multiple times to replace MASinGa clusters. The repetition of the same unique model can misrepresent the structural GC parameter distribution, biasing the simulated distribution toward the unique models' properties that were randomly chosen the most. To avoid such kind of bias, when determining  the radial distribution of each properties, only one unique model within each radial bin was considered. For each property, the mean value of each population's measurements has been determined together with the standard deviation.

\begin{figure}
    \centering
    \begin{subfigure}{0.5\textwidth}
       \centering
        \includegraphics[width=\linewidth]{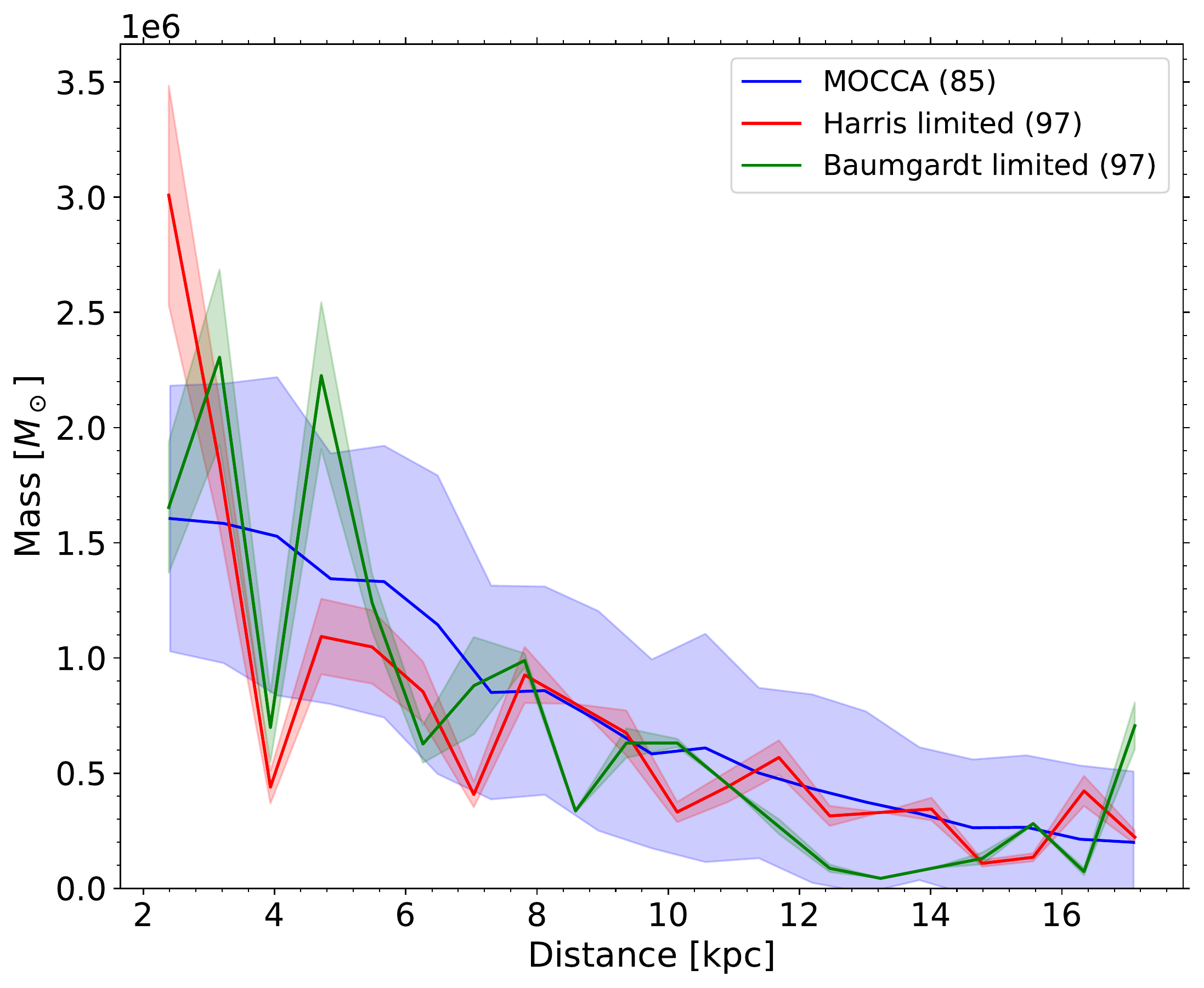}
    \end{subfigure}
    \begin{subfigure}{0.5\textwidth}
       \centering
        \includegraphics[width=\linewidth]{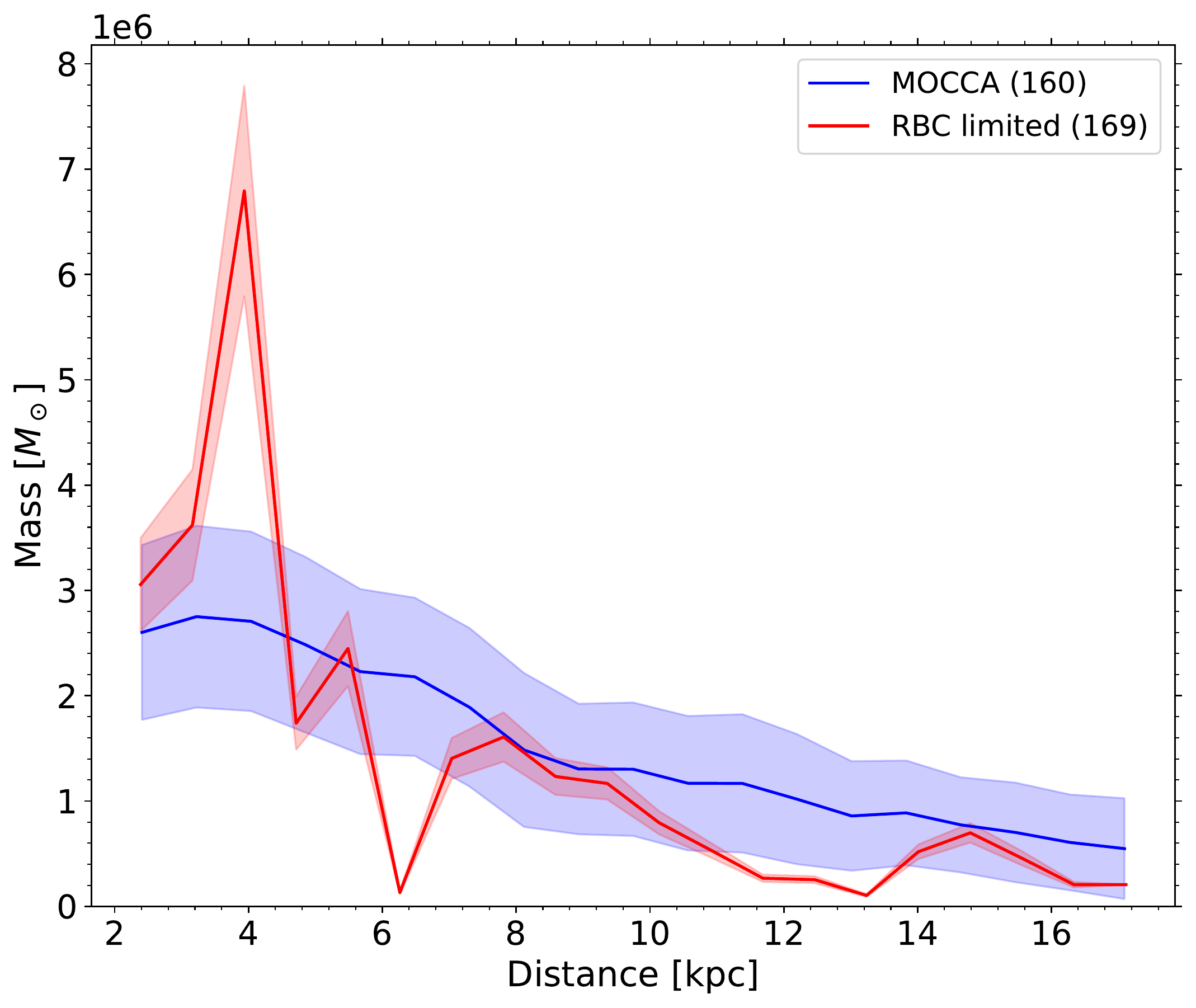}
    \end{subfigure}
    \caption{Mass distribution for the MOCCA population and the observed population for MW (top) and M31 (bottom) respectively. The shadow regions represent the standard deviations of the error for both the observed and the simulated GC populations. The mean number of surviving GCs are reported for MOCCA models, and the number of observed GCs are reported in parenthesis.}
    \label{Fig:massDesnityDistribution}
\end{figure}

\begin{figure}
    \centering
    \begin{subfigure}{0.5\textwidth}
       \centering
        \includegraphics[width=\linewidth]{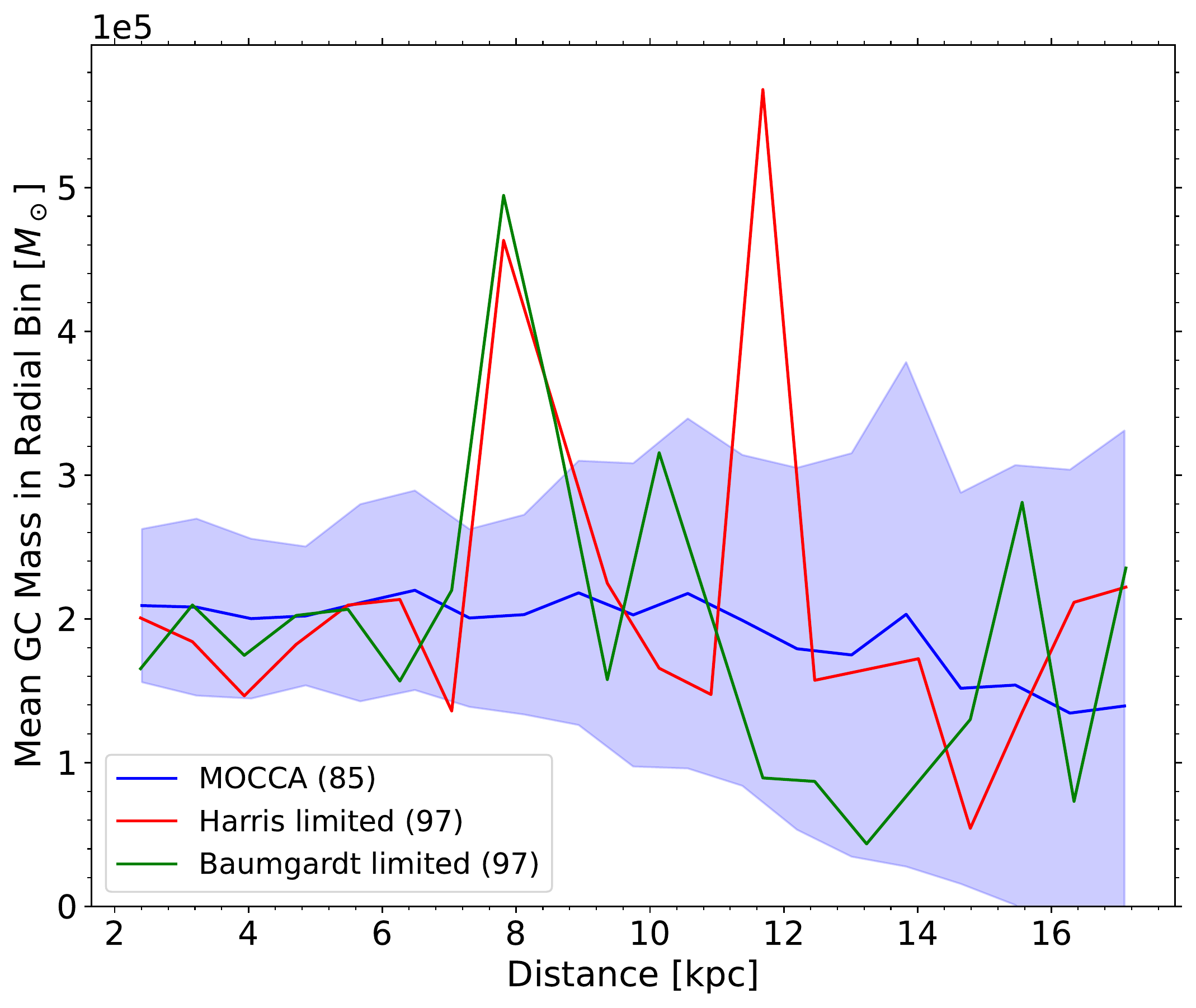}
    \end{subfigure}
    \begin{subfigure}{0.5\textwidth}
       \centering
        \includegraphics[width=\linewidth]{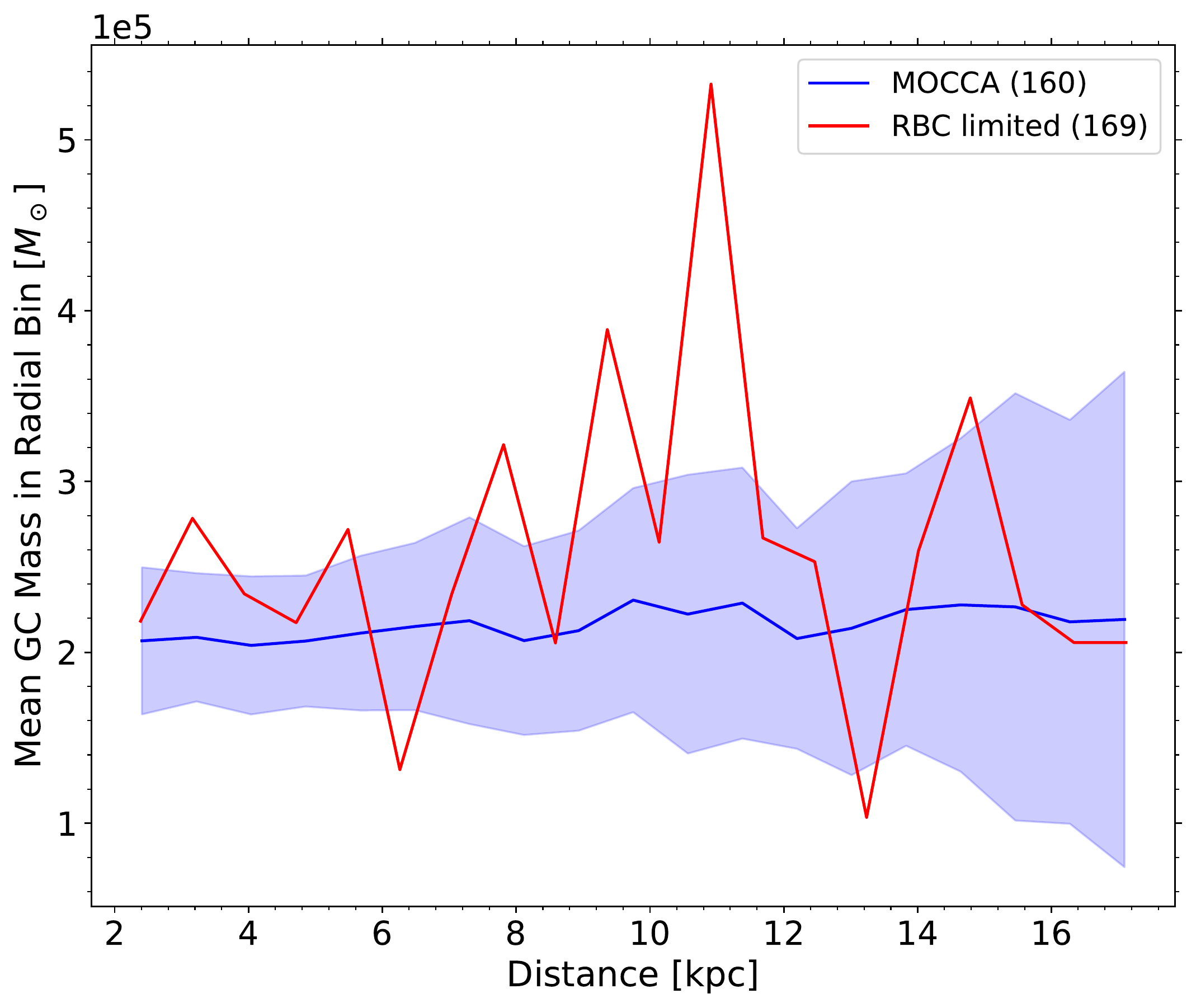}
    \end{subfigure}
    \caption{Mean mass distribution for the MOCCA population and the observed population for MW (top) and M31 (bottom) respectively.  The shadow regions represent the standard deviations of the error for the simulated GC populations. The mean number of surviving GCs are reported for MOCCA models, and the number of observed GCs are reported in parenthesis.}
    \label{Fig:meanMassDesnityDistribution}
\end{figure}

The mass distribution profiles of GC populations are reported in Fig. \ref{Fig:massDesnityDistribution}, for MW and M31. The GC populations have been divided into 20 galactocentric radial bins, and the mass  distribution has been determined as the total mass within the galactocentric distance bin. As it is possible to note, the GC's population is mostly concentrated within 5 kpc from the galaxy center. Also, the simulated distribution falls within the error limits of the observed distribution. However, a central steep increase is seen in the observational data, meanwhile our results do not show such prominent growth in comparison. As mentioned before, this difference can be due to the insufficiency of the MOCCA model to represent the region within 5 kpc from the galactic centre. On the other hand, the simulated mean mass within the galactocentric distance bin distributions is comparable with observations at all galactocentric distance bins for both MW and M31 with a constant value for different radial bins, as shown in Fig. \ref{Fig:meanMassDesnityDistribution}. These results gave additional evidence that the GCs' spatial density profiles obtained by our simulations follow a similar profile to the observed GCs in the MW and M31. Finally, the spikes in the observational profiles are caused by the small number of GCs found within the radial bin.

\begin{figure}
    \centering
    \begin{subfigure}{0.5\textwidth}
      \centering
        \includegraphics[width=\linewidth]{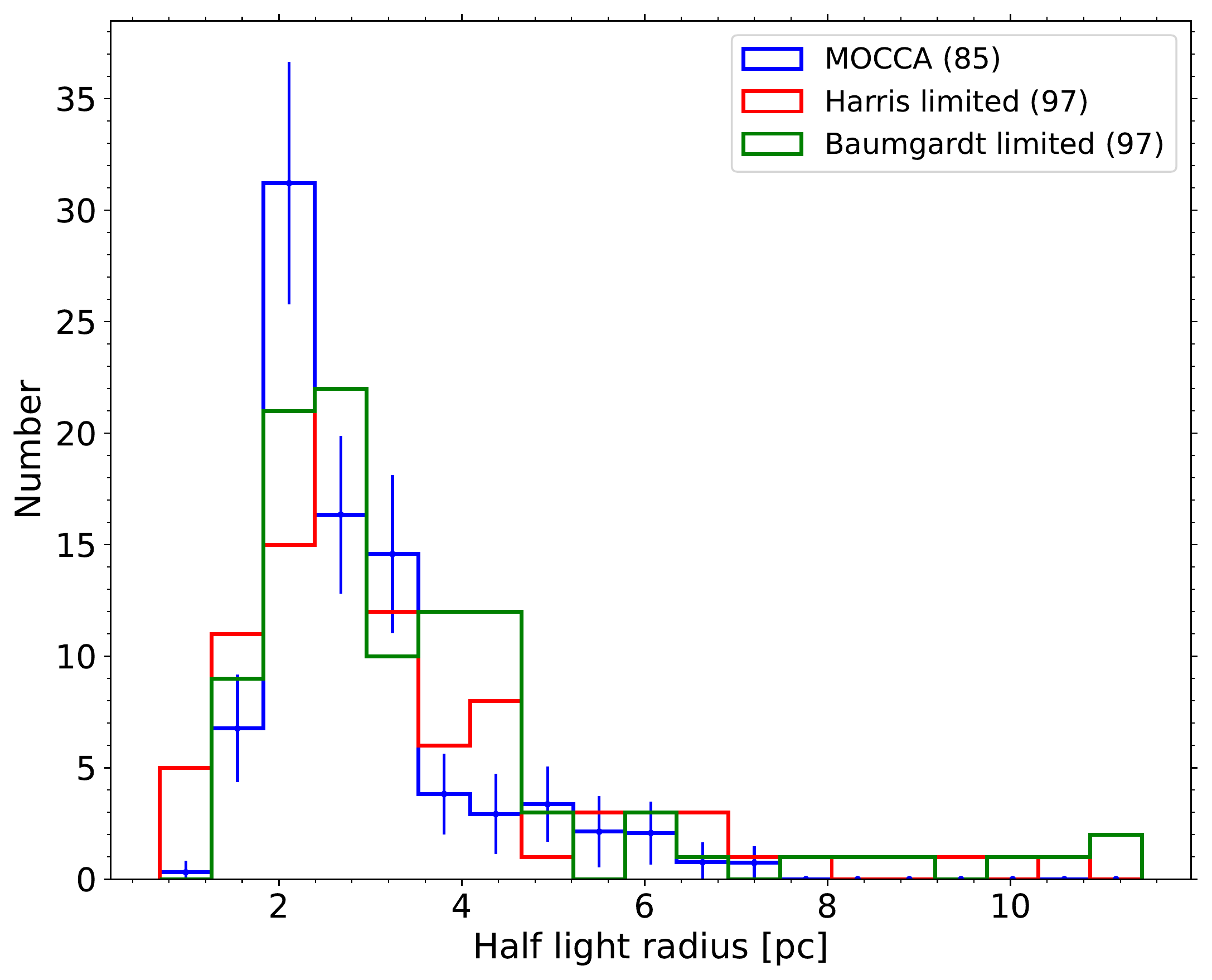}
    \end{subfigure}
    \begin{subfigure}{0.5\textwidth}
      \centering
        \includegraphics[width=\linewidth]{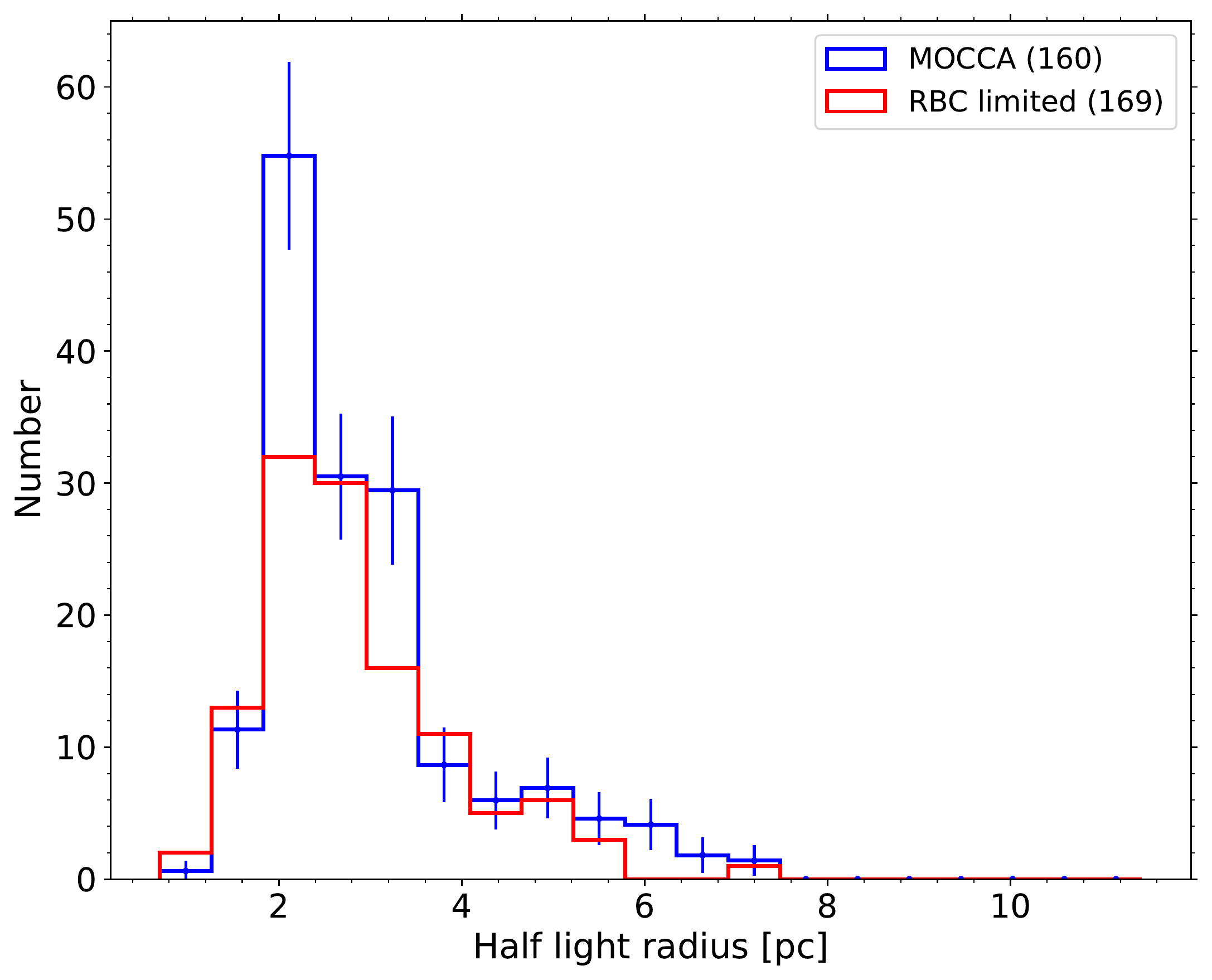}
    \end{subfigure}
    \caption{Half-light radius distribution for the MOCCA population and the observed population for MW (top) and M31 (bottom), respectively.  The mean number of surviving GC are reported for MOCCA models, and the number of observed GCs are reported in parenthesis. }
    \label{Fig:rhDistribution}
\end{figure}

The mean half-light radius distributions are properly reproduced, for both MW and M31, as it is possible to see in Fig. \ref{Fig:rhDistribution}. Most of the GC populations show a small half-light radius, with a peak around 2 pc for both MW and M31. As shown above, the GC populations are concentrated mostly in the central region of the galaxy, where the tidal field is stronger compared to the outer regions. The half-mass radius is expected to expand until the tidal field starts to control the system evolution. From that point the half-mass radius evolution would be regulated mainly by tidal mass loss. Additionally, at smaller galactocentric distances, the galactic density is higher, meaning a higher chance to be disrupted during close passages to the galactic center compared to the GCs in the outermost regions.Large GCs (with half-light radius greater than 8 pc) are not reproduced in our simulations, in contrast with the observed GCs in the MW. This is a consequence of the model selection described in Sec. \ref{sec:InitCond}, due to the half-light radius surface brightness limitation imposed on the  MOCCA models. 

In order to verify that our results are statistically in agreement with the the observed properties, we applied a two-sample Kolgomorov-Smirnov test (KS test) to those distributions. The KS test is used to compare two samples, quantifying the distance between the cumulative distribution functions, in order to verify the null hypothesis that the two samples are drawn from the same distribution. However, it is also possible to apply the alternative hypothesis, according to which the cumulative distribution of one sample is ``less'' or ``greater'' than the cumulative distribution of the other sample. In KS test terminology, a cumulative distribution that is ``greater'' than one other means that its mean and median will be smaller than the mean and median of the other distribution (vice-versa for ``smaller''). In fewer words, applying this alternative hypothesis means that the two distributions have the same shape, but the mean values are shifted, one with respect to the other. We applied also the alternative hypothesis (``less'' or ``greater'') to our sample, and a threshold value of $p \ge 0.05$. The best p-values and hypotheses (alternative and null) for the three comparison are showed in Table \ref{Table:KStest-results}. The reported best hypothesis indicate if our sample has a smaller mean (reported as ``Greater'') or greater mean (reported as ``Less'') compared to the observed samples. The results in Table  \ref{Table:KStest-results} show that our results are consistent with the observational data with a significance between $\sigma$ ($p \ge 0.8$) and $2 \sigma$ ($p \ge 0.17$), with the only exception being the  mass distribution for the Baumgardt catalogue comparison, with the best p-value close to the threshold acceptance criteria. The discrepancies seen for the Baumgardt catalogue comparison are connected to the spikes in the observational profiles, as explained before.

\begin{table*}
    \centering
    \begin{tabular}{ c@{\hskip 0.5in} cc cc cc}
    \hline
        & \multicolumn{2}{c}{Mass distribution} & \multicolumn{2}{c}{Mean mass distribution } & \multicolumn{2}{c}{Half-light radius}    \\
        \hline
        Catalogue & p-value & Hypothesis & p-value & Hypothesis & p-value & Hypothesis \\
        \hline
        Harris & 0.82 & Greater & 0.17 & Less & 0.95 & Less\\
        Baumgardt & 0.17 & Two-sided & 0.08 & Two-sided & 0.81 & Greater\\ 
        RBC &  0.64 & Greater & 0.64 & Less& 0.95 & Greater \\
    \hline
    \end{tabular}
    \caption{Best results of p-values from the Kolgomorov-Smirnov test between our simulations and the observed properties, and the corresponding hypotheses for mass distribution (left), mean mass distribution (centre) and half-light radius (right).}
    \label{Table:KStest-results}
\end{table*}

\subsection{NSC and central massive BH evolution}
In our simulations, models with galactocentric distances smaller than 10 pc were considered as accreted into the NSC.  Since the internal dynamics have been followed for the MOCCA Database model, it is possible to determine the mass of the IMBHs (if present in the cluster) that have also been accreted into the NSC for the MOCCA results. In our models, the SMBH mass build-up is driven by the build-up and merger of the IMBH hosted by the infallen GCs. The IMBH mass is not determined in the MASinGa code, and for this reason the SMBH mass was not estimated. This calculation involves all the models reproduced during the simulations, not only the ones above 2 kpc as done in the post-processing procedure. Also, during our simulations, only GCs with initial distance of 2.5 kpc merged at the center of the galaxy.

The NSC and SMBH masses from observations, MOCCA, and MASinGa are shown in Table \ref{Table:NSC_MW} and Table \ref{Table:NSC_M31}, together with the number and the mean mass of surviving GCs and the number and the total mass of IMBH sunk in the NSC. The observed mass of the NSC in the MW has a value of $1.8 \pm 0.3 \times10^7 M_\odot$ (with an half-light radius of 4 pc) \citep{Chatzopoulos2015}, meanwhile the SMBH at the center of the MW is $4.23 \pm 0.14 \times10^6 M_\odot$ \citep{Chatzopoulos2015}. Similarly, the NSC mass in M31 is $3.5 \pm 0.8 \times10^7 M_\odot$ \citep{Lauer1993,Kormendy2013,Georgiev2016} (with an half-light radius of $\sim 12$ pc \citep{Neumayer2020,Peng2002}), meanwhile the M31 SMBH has a mass of $\sim 1.1-2.3\times10^8 M_\odot$ \citep{Bender2005}. The NSC mass obtained in our simulations is smaller than the observed one by one order of magnitude. Similarly to the NSC mass, the total built-up mass for the SMBHs is on the order of $\sim 10^4 M_\odot$, much smaller than the observed SMBH masses. The authors in \cite{Takekawa2021} reported 5 IMBH candidates in the center of the MW, each of them having a mass $\gtrsim 10^4 M_\odot$. In our simulations, the mean number of accreted IMBHs is in mean $\sim 1-5$ for MW and $\sim 1-3$ for M31. 
Finally, the number of survived clusters in MOCCA is $86 \pm5$ and $164\pm7$ for MW and M31 respectively. These values are smaller than the observed number of cluster in MW and M31. The mean masses of surviving GCs in our simulations are in relatively well agreement with the observations.

\begin{figure}
    \centering
    \begin{subfigure}{0.5\textwidth}
      \centering
        \includegraphics[width=\linewidth]{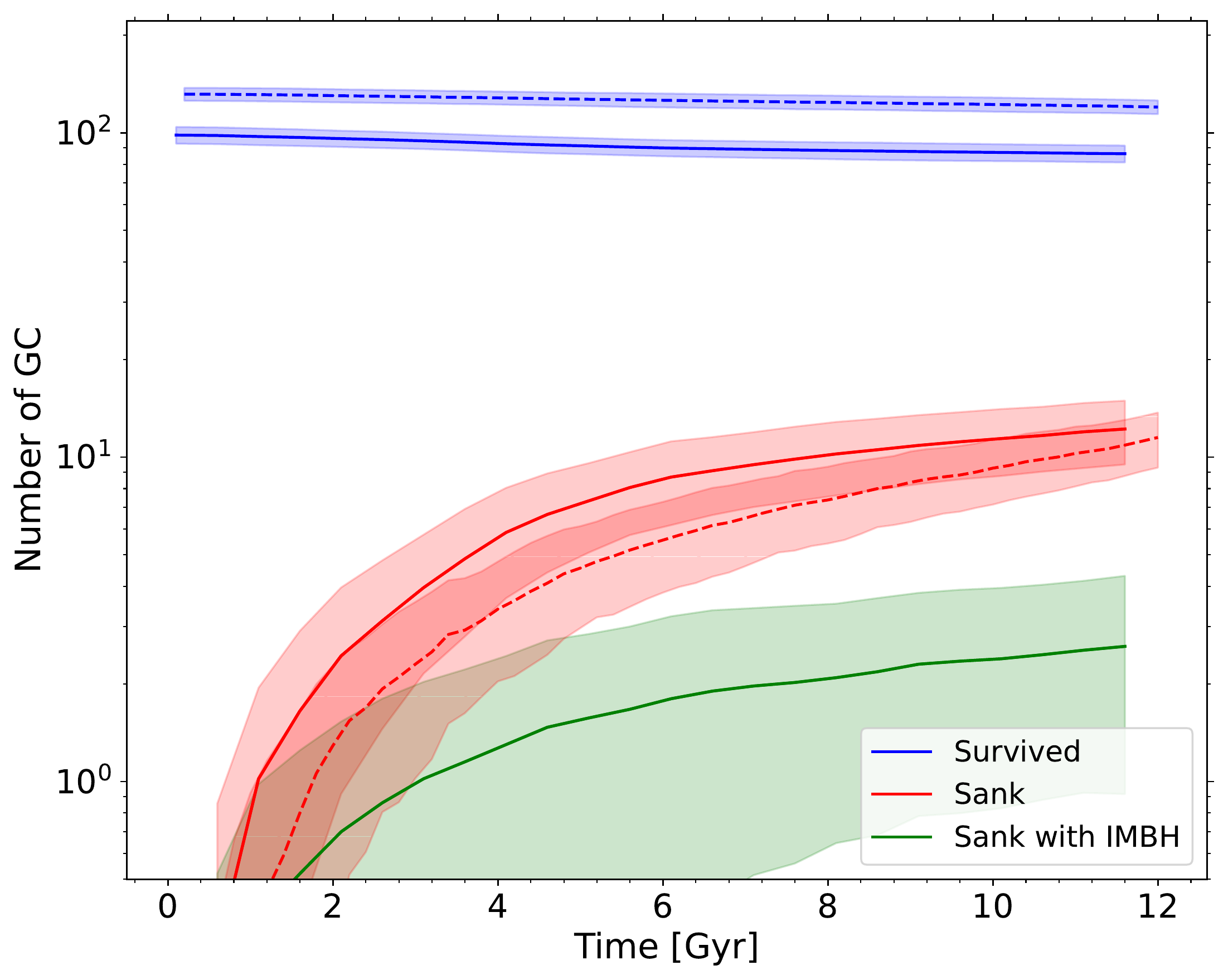}
    \end{subfigure}
    \begin{subfigure}{0.5\textwidth}
      \centering
        \includegraphics[width=\linewidth]{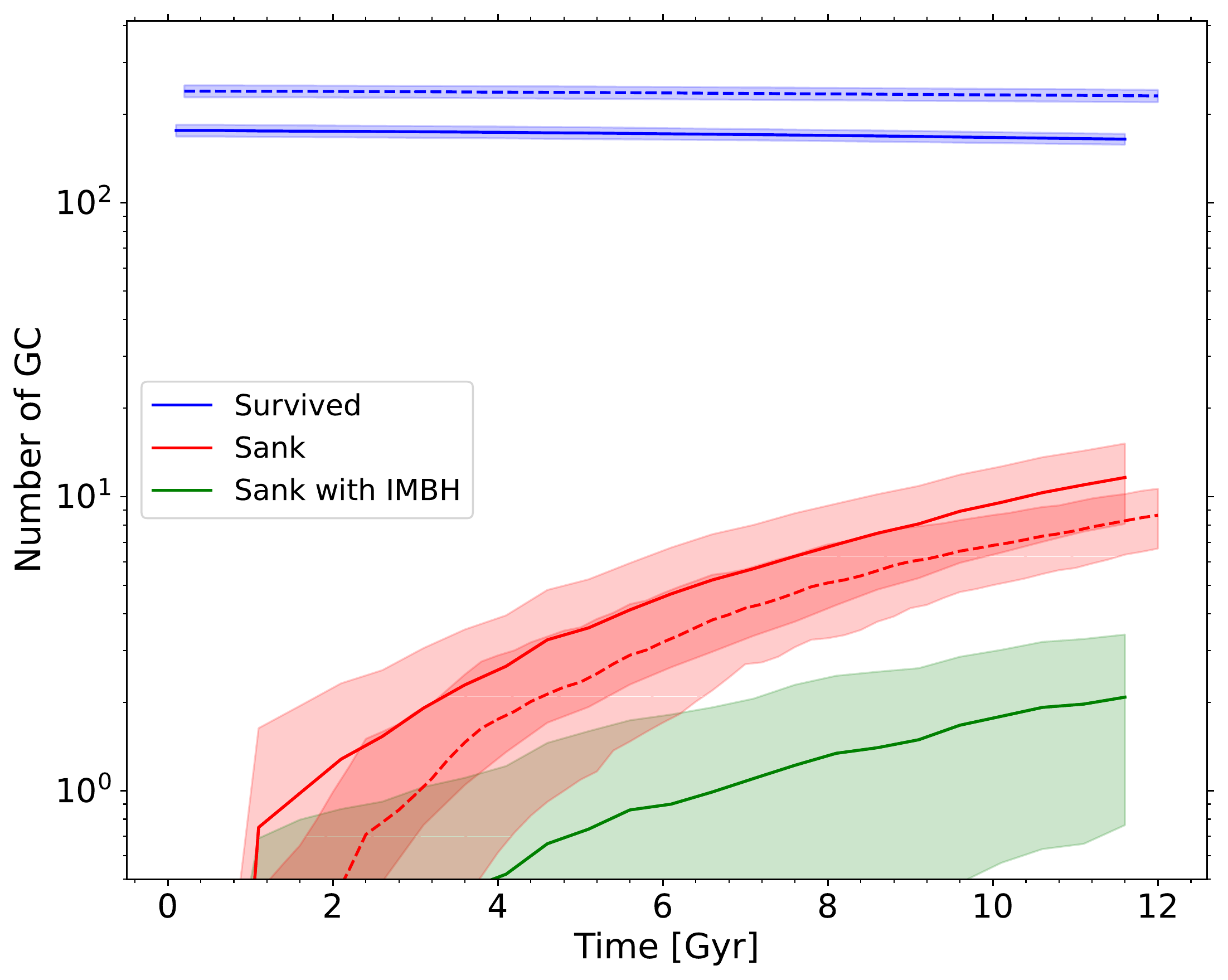}
    \end{subfigure}
    \caption{Time evolution of the mean number of GCs that survived (blue), that sank to the galactic centre (red), and that sank to the galactic centre hosting an IMBH (green) for MW (top) and M31 (bottom), respectively. The shadow region represents the standard deviation errors. The MOCCA and the MASinGa results are shown in solid and dashed lines, respectively.}
    \label{Fig:survived}
\end{figure}

As previously said, the MOCCA-Survey Database I does not reproduce properly GCs in the central region of the galaxy, influencing the final number and mass of GCs that would be accreted to the NSC or to the SMBH. In Fig. \ref{Fig:survived} we report the evolution in time for the number of GCs survived and sunk to the NSC, reporting also the number of sunk models hosting an IMBH. On average,  around 10\% for the MW and 5\% for M31 of the total initial GC populations sank into the NSC during the simulations, with only a very small percentage ($\sim 1\%$) of models hosting an IMBH that sank into the NSC. The self consistency of IMBHs in GCs and their accretion onto the NSC in our models are improvements with regard to previous works. These values do not change even when the galaxy's density in the  central regions ($<100$ pc) was increased to 10 times the actual galaxy's density. This simulated over-density would resemble the presence of a primordial NSC. Moreover, the rate of  infalling GCs in the galaxy centre is constant in time (apart from an important increase in the initial time) for M31 with a value of $3.2 \pm 1.2 \times10^5 M_\odot/Gyr$, meanwhile for the MW it was important in the first Gyr and it became less and less important at later times, with a value at 12 Gyr of $1.03 \pm 0.8 \times10^5 M_\odot/Gyr$.

\begin{table*}
\centering
\begin{tabular}{ccccccc}
Model & \# of GCs & Mean GC mass & NSC accreted mass & \# of IMBH in NSC & Total IMBH mass in NSC & Observed SMBH in NSC\\
\hline
Observations & 156 & $2.1 \pm 2.9 \times10^5$ & $1.8 \pm 0.3 \times10^7$& 5 & $\gtrsim 5\times10^4$ & $4.2 \pm 0.1 \times10^6$ \\ 
MOCCA & $86\pm5$ & $2.1 \pm 1.5 \times10^5$ & $3.4 \pm 1.0 \times10^6$& $3\pm2$ & $3.6 \pm 2.7 \times10^4$  & -\\ 
MASinGa & $120\pm6$ & $2.2 \pm 1.3 \times10^5$ & $3.3 \pm 0.9 \times10^6$&  - & - & -\\ 
\hline\end{tabular}
\caption{The number and the mean masses of survived GC, the NSC accreted mass, the number and the total mass of IMBHs accreted to the NSC and the observed SMBH mass from observations, MOCCA and MASinGa for MW. The mass values are in solar units. The values from the Harris catalogue have been used to determine the properties of the survived GCs.}
    \label{Table:NSC_MW}
\end{table*}

\begin{table*}
\centering
\begin{tabular}{ccccccc}
Model & \# of GCs & Mean GC mass & NSC accreted mass & \# of IMBH in NSC & Total IMBH mass in NSC & Observed SMBH in NSC\\
\hline
Observations & 231 & $4.7 \pm 2.0 \times10^5$ & $3.5 \pm 0.7 \times10^7 $& -& - & $\sim 1.1-2.3\times10^8 $ \\ 
MOCCA & $164\pm7$ & $2.1 \pm 1.4 \times10^5$ & $3.3 \pm 1.2 \times10^6$& $2\pm1$ & $2.7 \pm 2.0 \times10^6$& - \\ 
MASinGa & $231\pm11$ & $2.3 \pm 1.3 \times10^5$ & $3.1 \pm 1.0 \times10^6$&  - & -& -\\ 
\hline
\end{tabular}
\caption{The number and the mean masses of survived GC, the NSC accreted mass, the number and the total mass of IMBHs accreted to the NSC and the observed SMBH mass from observations, MOCCA and MASinGa for M31. The mass values are in solar units.}
    \label{Table:NSC_M31}
\end{table*}

\section{Discussion} \label{sec:Discussion}
The model evolution recipe used in MASinGa has been carried out with a few simplifications for the half-mass radius, the tidal radius and the mass evolution. With simplistic physical assumptions, the equations used to describe their evolution have been determined. The galaxy density profile has been described by a Dehnen model \citep{Dehnen1993}, with the assumption that the initial density profile for the galaxy is similar to the currently observed ones. 

The properties of our simulated GC populations are in agreement with the observed properties for both the MW and M31, despite the simplifying assumptions and limitations of our models. Our simulations show a large density distribution of models in the central region of the galaxy ($<5$ kpc),  with a decreasing density at larger galactocentric distances. A similar trend is seen in the observed population. Therefore, it is expected that the GC populations would be composed mostly of compact GCs, with half-light radii on the order of few pc, as seen in both observations and simulations. Indeed, the interplay of a smaller tidal radius and larger galaxy density, would not allow the GCs to expand substantially since they would be disrupted by interaction events with the galaxy. To quantify the quality of our results, a  two-sample Kolgomorov-Smirnov test was applied to the observed and simulated distributions, for different alternative hypotheses. The results show that the simulated and observed distributions likely come from the same distribution, with a significance between $\sigma$ and $2\sigma$.

The models from the MOCCA Database I have been used to reproduce the MW and M31 GC populations. A non uniform initial mass distribution in the MOCCA Dabase I models could put some limitations on the reproduction of the initial GCIMF and observed final masses. Moreover, as reported in \cite{Madrid2017}, the MOCCA results were able to reproduce the N-body simulations for Galactocentric distance down to few kpc. Considering also an under-estimation of galaxy density and mass in the central region due to the limitations of the Dehnen model, the number and the evolution of GCs in the central regions could have been under-estimated. 

\begin{figure}
    \centering
        \includegraphics[width=\linewidth]{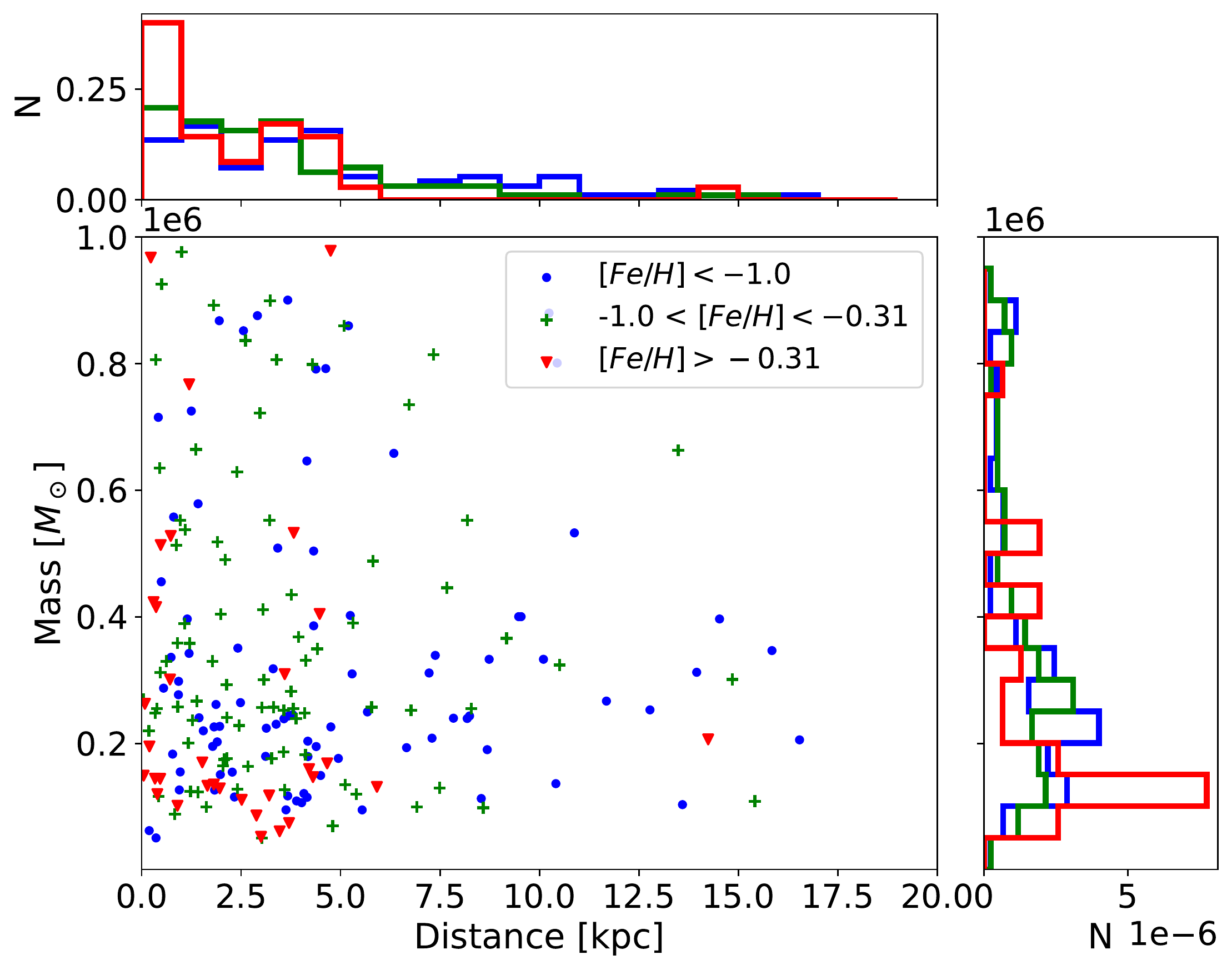}
    \caption{Observed masses versus the projected position for the M31 GCs, for different metallicities. In red, in green, and in blue the metal-rich, the intermediate metallicity, and metal-poor GCs are shown respectively. On the side, the normalized histogram showing the distribution of each population is reported.}
    \label{Fig:metallicity}
\end{figure}

One additional source of differences seen in our simulations and the observed GC populations can be the bimodal nature of GCs. It is known that both the MW and M31 present two GCs populations: a blue, metal-poor one and a red, metal-rich one. Generally, the metal rich red clusters are expected to form during the gas-rich mergers during minor galaxies mergers. On the other hand, the blue GCs would be formed in the progenitor galaxies. For this reason, it is expected that the metal-rich GCs would be mostly centrally concentrated, and the metal-poor ones more spatially redistributed during the galaxies' collisions \citep{Renaud2017}. This is indeed observed in both MW and M31 populations, as  is possible to see in Fig. \ref{Fig:metallicity}. The figure shows the observed mass and projected positions for GCs in M31 for different metallicities. The GCs have been divided into three groups depending on their metallicities: metal-poor ones with $[Fe/H] < -1.0$, metal-rich with $[Fe/H] > -0.31$ and intermediate metallicity GCs with $-1.0<[Fe/H] < -0.31$. The observed over-density in the central regions is actually predominated by the presence of metal-rich and intermediate metallicity GCs, even though the metal-rich GCs represent a small percentage of the total population. Similar results are obtained for the MW GC population. On the other hand, our simulations assumed that all GCs were generated simultaneously at the initial times. This means that our models do not take into account the presence of different GC populations.

The number of infalling GCs seems to be a small percentage of the total GC population, being around 10\% and 5\% for the MW and M31 respectively, with an even smaller percentage for infalling GCs hosting an IMBH ($\sim 1\%$). The rate of infalling events seems to be constant for the M31 model evolution, with a small peak in the initial times, whereas the event rate for the MW seems to be more important at later times. As mentioned earlier, the central galactic region is not properly reproduced by the MOCCA  models.

On the other hand, as shown in the Appendix \ref{Appendix}, the polynomial fit to the rotational curve improves the estimated galaxy mass in the central region, as seen in observations. Meanwhile, its overall GC population's properties are not different from the Dehen density fit model with a larger number of infalling GCs having been observed together with a larger NSC and SMBH masses. The NSC and SMBH do show larger final build-up masses in the polynomial fit model compared to the rotational curve. However, the final masses are still importantly smaller than the observed values, meaning that the final NSC and SMBH masses are not strongly influenced by a better representation of the galaxy density in the central region. Given the flexibility that the Dehnen model allows, it will be possible to populate GCs around a larger number of external galaxies easily and automatically using the Dehnen family models. As shown the galaxy density profile represented by a Dehnen model can actually reproduce adequately the observed GCs population properties. 

The formation mechanics of NSCs is still an open topic, with two main mechanics principally considered: the \textit{in situ} formation, and the GC infall and merging with the galactic centre. In the \textit{in situ} formation scenario,  the pristine gas would fall into the galactic centre and boost an intense burst of star formation \citep{Loose1982,Milosavljevic2001,Bekki2007,Neumayer2011}. In the second scenario, the dynamical friction would slowly spiral the GCs inwards, and eventually they would be accreted in the galaxy centre \citep{Tremaine1975,CapuzzoDolcetta1993,CapuzzoDolcetta2009,ArcaSedda2014}. Finally, \cite{Guillard2016} showed that the interplay of those two mechanics may explain the formation and evolution of NSC, with a massive star cluster formed in the disc of the galaxy, migrating to the centre and increasing its mass through interactions with other star clusters and substructures. The final NSC mass from our simulations is roughly 10\% of the observed NSC mass in both MW and M31, meaning that the mechanics for the NSC build up mass is driven not only by the infalling scenario, but by the initial accreted mass and mergers with the fallen GCs operating together \citep{Urry1995,vandenBosch2012,Kormendy2013,Emsellem2013}. 

Single SMBHs are often hosted at the centres of galaxies, with masses ranging $10^6 - 10^{10} M_\odot$. For galaxies with masses between $10^{10} - 10^{11} M_\odot$ the co-existence of a SMBH and NSC is observed \citep{Seth2008,Leigh2012,Scott2013}. Galaxies with masses below $10^{10}$ or above $10^{11} M_\odot$ are dominated by the presence of either a NSC or a SMBH, respectively. The observed scaling relations between the host galaxy, the NSC and the SMBH suggests a continuous sequence of NSC- and SMBH- dominated galaxies \citep{Bekki2010}. The main proposed scenario to explain the formation of SMBHs is the formation and the merger of massive stellar remnants which sink into the galaxy centers \citep{Quinlan1990,Volonteri2005,Ebisuzaki2001}. One other possible formation scenario of a SMBH is the collapse of super-massive primordial gas and the evolution the super-massive object which forms as a result \citep{Haehnelt1993,Gnedin2001,Bromm2003}.

Recent works show that the \textit{in situ} gas growth and mergers of young stellar cluster that formed nearby in the Galactic center can contribute importantly to the NSC and SMBH mass growth. Using direct N-body simulations to reproduce the merge of stellar clusters and using a very simplified growth model for the SMBH and NSC masses, the authors in \citep{Askar2021,Askar2021-2} found that for galaxies like the MW (with stellar masses close to $10^{10}$ and $10^{11} \,\,M_\odot$), the  gas growth can be very important in increasing both the NSC mass and as well as the SMBH mass. Also, about $10 - 15\%$ of the stars that compose the mass of the NSC in the MW are actually old metal poor stars, with abundances that are similar to the ones observed in the GCs \citep{Do2020,ArcaSedda2020}. These results are in agreement with the value reported in this paper, with around $10\%$ of the NSC mass being explained by the dry merger scenario. In fact, we applied the procedure described in \citep{Askar2021,Askar2021-2} to our MW and M31 results. We assumed that the IMBH delivered to the NSC may accrete the gas in the central region of the galaxy before the delivery of a next IMBH, with a 10\% of Eddington accretion rate during this phase. This means that the IMBH mass would double in a time-scale of $\sim 300$ Myr, and in our calculation, a random value between 250 and 350 Myr was used. We applied this calculation only if the IMBH was delivered before 4.5 Gyr. Finally, not all the gas present in the galactic center could be accreted in the SMBH. The remaining gas can contribute to the \textit{in situ} star formation in the NSC, and eventually induce the NSC mass growth. The contribution of this star formation event is related to the final SMBH mass, and it was randomly chosen between 0.8 and 3 for MW and 0.8 and 1.5 for M31 respectively. Using these simple and ad-hoc prescriptions, we applied this procedure to all the 100 representation of GC population for both MW and M31, finding that the SMBH and the NSC final masses were growing by few order of magnitudes, with values comparable to the observations. A more detailed study will be performed in the future works.

\section{Conclusion}
In this paper we introduced the machinery that will be used in the next works to populate the local Universe galaxies GC populations with the MOCCA models. The reproduction of the MW and M31 GC populations has been carried out using the semi-analytic modelling code MASinGa \citep{ArcaSedda2022}. The MASinGa code has been updated and extended, with the internal dynamical evolution described by the MOCCA-Survey I Database models instead of the analytic approximations.

The MW and M31 have been populated with 100 GC population representations, evolving them up to 12 Gyr. The mean properties obtained from these representation have been compared to the observed GC populations' proprieties. The results shown are in agreement with the observed properties for both the MW's and M31's GC populations. Similarly, the NSC and SMBH masses found in our models are in agreement with the  dry merger scenario.

Summarizing our main results:
\begin{itemize}
    \item The spatial distributions for the MW's and M31's GC populations have been reproduced, with a large amount of the population observed within a galactocentric distance of 5 kpc, as shown in Fig. \ref{Fig:densityMapMW} and Fig. \ref{Fig:densityMapM31}. The observed mass profile of the GC populations also shows an important increase in the central region of the galaxy, not reproduced by our simulations - see Fig. \ref{Fig:massDesnityDistribution}.
    \item In the central galactic regions, the stronger tidal field and higher galactic density would constrain the GCs expansion and mass loss, implying that only dense and compacted GCs would survive the galaxy interactions. As a result, most of the GCs are relatively compact and have a half mass radius smaller than 4 pc, as shown in Fig. \ref{Fig:rhDistribution} for both observations and simulations, in the MW and M31.
    \item The GCs' galactocentric distance evolution has been followed down to 10 pc, with GCs considered accreted to the NSC for smaller distances. The mass accretion rate in the galactic centre seems to be constant in time, with values of $\sim 1-3\times10^5\,\,M_\odot/Gyr$ for both the MW and M31.
    \item The SMBH mass build-up has been considered as the accretion of GCs hosting IMBHs that have fallen into the NSC during the simulation - see Table \ref{Table:NSC_MW} and Table \ref{Table:NSC_M31}. The final NSC and SMBH masses determined by our simulations are smaller than the observed values by few order of magnitudes. These differences do show that the NSC and SMBH mass build-up cannot be explained completely and only by infalling scenario model, and that the interplay of the formation on an initial accreted mass and the interactions and merges with infalling GCs is needed. 
\end{itemize}

Our work lays the ground for a series of future explorations which will focus on the impact of galaxy-GC co-evolution on the formation of compact object binaries, IMBHS, and GW sources. We aim to constrain and determine not only the GCs' observational properties, evolutionary paths, and their compact object content (such as IMBH, BHS, BH-BH binaries, X-ray binaries), but also the NSC and the central SMBH mass build-up. The results from our simulations could be used to determine the BH-BH merger rate in the local Universe, together with the  event rates of TDEs between the SMBH and the infalling GCs. 

\section*{Acknowledgements}
MG and AL were partially supported by the Polish National Science Center (NCN) through the grant UMO-2016/23/B/ST9/02732. MAS acknowledges financial support from the European Union’s Horizon 2020 research and innovation programme under the Marie Skłodowska-Curie grant agreement No. 101025436 (project GRACE-BH, PI Manuel Arca Sedda). AA acknowledges support from the Swedish Research Council through the grant 2017-04217.

\section*{Data Availability}
The data underlying this article will be shared on reasonable request to the corresponding author.
\bibliographystyle{mnras}
\bibliography{biblio}

\appendix

\section{Polynomial fit to the rotational curve M31 results} \label[Appendix A]{Appendix}
As discussed in Sec. \ref{subSec:initCond}, the results shown so far are obtained from the Dehnen model fit for the observed rotational velocity curve. This fit, as shown in Fig. \ref{Fig:rotCurve}, does not properly reproduce the M31 central region rotational velocity. A polynomial curve was fitted to the observed rotational velocity, in order to better estimate the M31 central mass. The Dehnen best fit parameters were used to estimate the dynamical friction. Instead, the mass and density distribution, together with the GCs' density distributions are determined using the polynomial curve. In this way, the central GCs' density distributions are enhanced compared to the Dehnen model fit.

\begin{figure}
    \centering
      \centering
    \includegraphics[width=\linewidth]{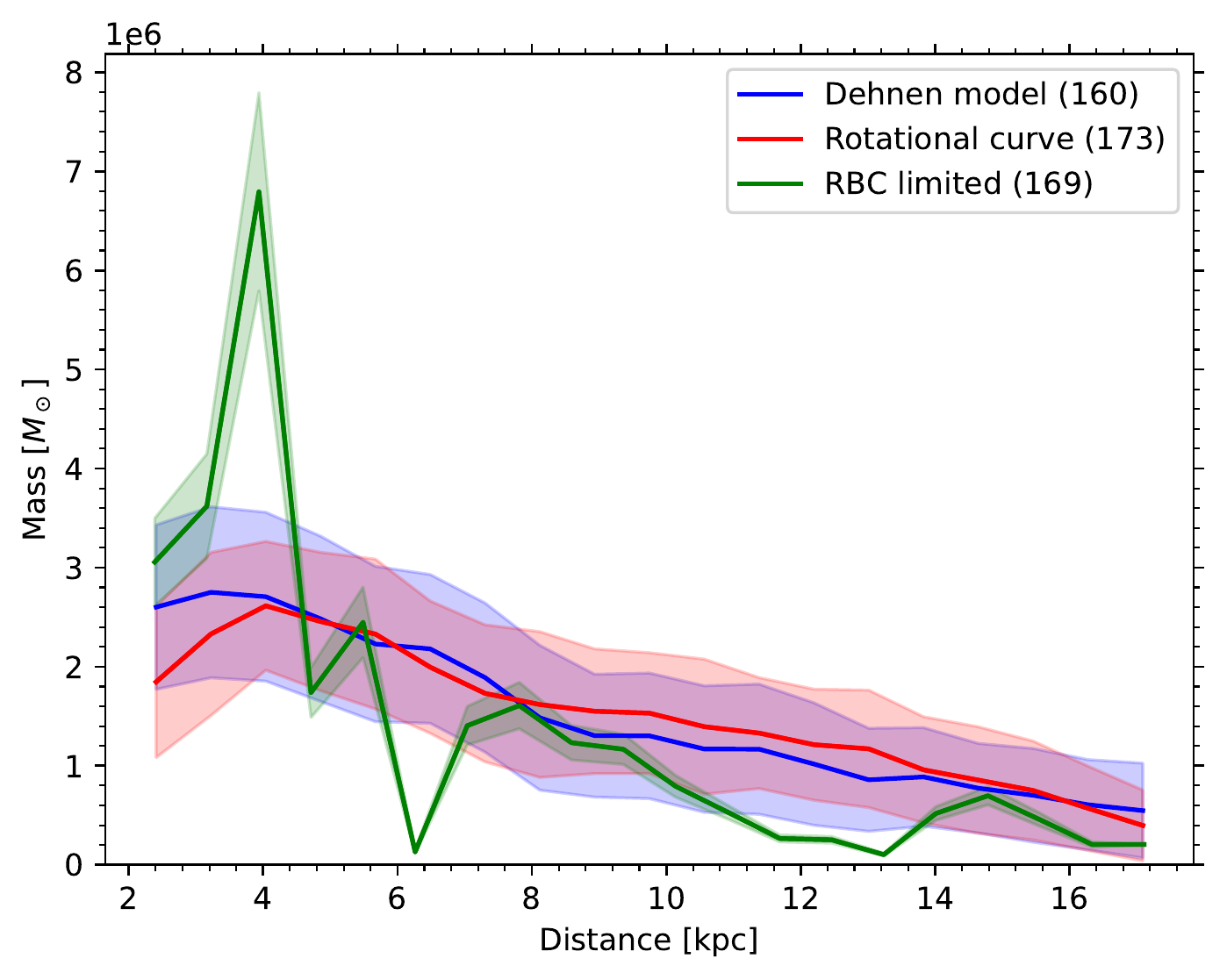}
    \caption{Mass distribution for the MOCCA population and the observed population for M31, from the models with the Dehnen model (blue) and the polynomial curve fit (red). The shadow regions represent the standard deviation error for both the observed and the simulated GC populations. The mean number of surviving GCs are reported for MOCCA models, and the number of observed GCs are reported in parenthesis.}
    \label{Fig:massDesnityDistributionRot}
\end{figure}

Even though the central region was more populated (with more GCs in number and in mass) in the initial condition, the over-density of GCs in the central region seen in observations is still not reproduced.  In Fig. \ref{Fig:massDesnityDistributionRot} the mass distribution obtained with this model fit is shown, compared to the mass distribution obtained from the Dehnen model fit. The results do not differ much from the ones obtained from the Dehnen model fit only (see Fig. \ref{Fig:massDesnityDistribution}), indicating that the over-density seen in  observations is not only related to the not precise rotational curve fit and galaxy mass distribution in the central region. Similarly, the mean GC mass distribution and the half-light radius distribution show similar results to the the Dehnen model fit. This means that the GC's surviving population's properties is not strongly affected by a better rotational curve model representation in the central region of the galaxy. 

On the other hand, comparing the value reported in Table \ref{Table:NSC_M31} and in Table \ref{Table:NSC_M31_rot}, a larger NSC mass is obtained from a polynomial fit to the rotational curve compared to the Dehnen fit model, with the final SMBH mass being 2 times larger than the  Dehnen fit model value. However,  these values are still largely smaller than the observed values. Instead, a larger number of infalling GCs have been observed in the polynomial fit simulation. Overall, a better fit to the rotational curve velocity for M31 does not influence the surviving GC population properties. On the other hand, a slight increase of the NSC and SMBH masses is observed, but it is still not so important as to be comparable with observations. These  results show that the Dehnen density model is adequate enough to describe the galaxy density profile. 

\begin{table*}
\centering
\begin{tabular}{ccccccc}
Model & \# of GCs & Mean GC mass & NSC accreted mass & \# of IMBH in NSC & Total IMBH mass in NSC & Observed SMBH in NSC\\
\hline
\hline
Observations & 231 & $4.7 \pm 5.0 \times10^5$ & $3.5 \pm 0.7 \times10^7 $& - & -& $\sim 1.1-2.3\times10^8 $ \\ 
MOCCA & $177\pm8$ & $1.8 \pm 1.4 \times10^5$ & $4.7 \pm 1.0 \times10^6$& $4\pm2$ & $6.4 \pm 3.0 \times10^6$& - \\ 
MASinGa & $261\pm11$ & $2.1 \pm 1.3 \times10^5$ & $2.2 \pm 0.9 \times10^6$&  - & -& -\\ 
\hline\end{tabular}
\caption{The number and the mean masses of survived GC, the NSC accreted mass, the number and the total mass of IMBHs accreted to the NSC and the observed SMBH mass from observations, MOCCA and MASinGa for the M31 polynomial fit result. The mass values are in solar units.}
    \label{Table:NSC_M31_rot}
\end{table*}

\bsp
\label{lastpage}

\end{document}